\documentclass[%
reprint,
superscriptaddress,
amsmath,amssymb,
aps,
floatfix,
]{revtex4-1}
\usepackage{graphicx}
\usepackage{upgreek}
\usepackage[english]{babel}

\begin{document}
\newcommand{\tb}[1]{{\color{red}#1}}
\newcommand{\ket}[1]{|#1\rangle}
\newcommand{\bra}[1]{\langle#1|}
\newcommand{\braket}[2]{\langle#1|#2\rangle}
\newcommand{\ovlap}[2]{|\braket{#1}{#2}|^2}
\newcommand{\proj}[1]{{|#1\rangle\!\langle#1|}}


\title{Quantum detector tomography of a 2$\times$2 multi-pixel array of superconducting nanowire single photon detectors}

\author{Timon Schapeler}\email{timons@mail.upb.de}


\author{Jan Philipp H\"opker}
\author{Tim J. Bartley}
\affiliation{Mesoscopic Quantum Optics, University of Paderborn, Warburger Stra\ss e 100, 33098 Paderborn, Germany}




\begin{abstract}
We demonstrate quantum detector tomography of a commercial 2$\times$2 array of superconducting nanowire single photon detectors. We show that detector-specific figures of merit including efficiency, dark-count and cross-talk probabilities can be directly extracted, without recourse to the underlying detector physics. 
These figures of merit are directly identified from just four elements of the reconstructed positive operator valued measure (POVM) of the device. We show that the values for efficiency and dark-count probability extracted by detector tomography show excellent agreement with independent measurements of these quantities, and we provide an intuitive operational definition for cross-talk probability. 
Finally, we show that parameters required for the reconstruction must be carefully chosen to avoid oversmoothing the data.
\end{abstract}

\maketitle

\section{Introduction}
Due to their low noise, high timing resolution and excellent efficiency across a broad wavelength range, superconducting nanowire single photon detectors (SNSPDs) have become increasingly prevalent in low-photon-flux optical sensing~\cite{gol2001picosecond,natarajan_superconducting_2012,marsili_detecting_2013,esmaeil_zadeh_single-photon_2017,korzh2020demonstration}. Advances in fabrication yield 
and read-out techniques~\cite{mccaughan_readout_2018,miyajima2018high,cahall2018scalable,gaggero2019amplitude,tiedau2020single,allmaras2020demonstration,takeuchi2020scalable} have enabled arrays of such detectors to be developed~\cite{dauler_multi-element_2007,divochiy_superconducting_2008,marsili_physics_2009,jahanmirinejad_photon-number_2012,zhao_superconducting-nanowire_2013,rosenberg_high-speed_2013,verma_four-pixel_2014,allman_near-infrared_2015,najafi_-chip_2015,chen_16-pixel_2018,tao2019high,wollman2019kilopixel}, which have lead to applications in imaging~\cite{miki_64-pixel_2014,allman_near-infrared_2015,moreau2019imaging} and deep space communication~\cite{shaw_arrays_2015}. Such devices are also ideal for multiplexed photon counting in quantum optics experiments. Photon counting with SNSPDs can be achieved in multiple ways, either through the intrinsic response of the device~\cite{cahall2017multi,zhu2018scalable,zhu2020resolving,zou2020superconducting} or by spatial or temporal multiplexing~\cite{natarajan2013quantum,tiedau2019high}.

In order to use such detectors effectively, particularly in the field of quantum optical technologies, a quantum mechanical description of these devices is necessary. Typically, a ``bottom-up'' approach is employed, based on modelling the underlying physics governing the working principles of such detectors
. Detector parameters such as efficiency and noise are put into these models \textit{a priori}. On the other hand, ''top-down'' techniques such as quantum detector tomography can provide an operational description of the device, including many of the figures of merit such as efficiency, dark-count and cross-talk probabilities, without recourse to an underlying model of the detector's working principle, geometry or readout scheme. 

Quantum detector tomography concerns finding the elements of the so-called positive operator valued measures (POVMs), which fully characterises the detection operation. In the context of photon counting, these elements represent probabilities of different outcomes of the detector, given a particular number of incident photons. This treats the detector itself as a black box, which is characterised by known inputs and measured outputs. Quantum detector tomography has been applied to several different single-photon detectors, including avalanche photodiodes~\cite{lundeen2009tomography,feito2009measuring,piacentini2015positive} and SNSPDs~\cite{natarajan2013quantum}, in both single-channel and time-multiplexed geometries. It has also been applied to transition-edge sensors, which resolve energy at the single-photon level~\cite{humphreys2015tomography}, as well as coherent detection schemes~\cite{zhang2012mapping,cooper2014local}.
Spatial arrays of single-photon detectors have been studied in a ``bottom-up'' approach, based on modelling their operating principle~\cite{afek2009quantum,nehra2020photon}. However, ``top-down'' detector tomography 
has not, to the best of our knowledge been carried out on spatial arrays. 
Such arrays are important since they exhibit additional noise sources such as cross-talk, which is not only essential to characterise for accurate measurements, but also challenging to measure as the size of the array increases~\cite{eraerds2007sipm,akiba2008multipixel,afek2009quantum,kalashnikov2012crosstalk,nehra2020photon}. Furthermore, from a fundamental perspective, they represent quantum objects spanning an extremely large Hilbert space, and are among the largest objects to be described in a fully quantum-mechanical manner. It is therefore necessary to demonstrate that quantum detector tomography works in principle for such arrays, and that the techniques can be scaled with their size. 

In this paper, we address the first of these issues, by presenting a tomographic reconstruction of the response of a 2$\times$2 array of SNSPD pixels. Using this technique, we can directly quantify the effects of cross-talk, as well as distinguish this from dark noise and determine the detector efficiency. Whilst the salient physics is present in the 2$\times$2-array, this method can be readily generalised to  much larger arrays, where characterising pixel-by-pixel becomes highly challenging (and characterising the interpixel correlations prohibitive) with increasing array size.

\section{Detector Tomography}
As introduced by Lundeen et al.~\cite{lundeen2009tomography}, the aim of quantum detector tomography is to reconstruct the set of positive operator valued measure (POVM) elements $\left\{\pi_n\right\}$. These elements fully describe the action of the detector during a measurement, namely by relating the set of outcomes $p_{\rho,n}$ to the set of input states $\left\{\rho\right\}$ through the equation 
\begin{equation}\label{eqn:Born}
 p_{\rho,n}=\mathrm{Tr}\left[\rho \pi_n\right]~.   
\end{equation}
If the input states $\left\{\rho\right\}$ and outcome statistics $p_{\rho,n}$ are known, then this equation can be inverted to find the POVM set $\left\{\pi_n\right\}$, under the positive semidefinite constraints $\pi_n>0$ and $\sum_{n}\pi_n=1$, in order to represent physically meaningful probabilities. 

The choice of input states is in principle arbitrary, with the sole requirement that they span the same Hilbert space as the detector. Coherent states are ideal candidates since they are both overcomplete and straightforward to produce in the laboratory. Coherent states are fully characterised by their mean photon number, which must be precisely determined prior to characterising the detector under test. While a single coherent state is in principle sufficient to span the Hilbert space of the detector, it is also important to obtain enough outcome statistics across the whole Hilbert space in a reasonable measurement time, therefore a set of coherent states with different mean photons numbers are chosen. 

Under the assumption that the detector is insensitive to the phase of the incoming light field, its POVMs contain only diagonal elements, {i.e.}
\begin{equation}\label{eqn:pin}
    \pi_n=\sum_{i=0}^\infty\theta^{(n)}_i\proj{i}~,
\end{equation}
where $\theta^{(n)}_i$ represents the probability of outcome $n$ given $i$ incident photons. Using this notation, Eq.~(\ref{eqn:Born}) can be recast as the matrix equation
\begin{equation}
\mathbf{P}=\mathbf{F}\mathbf{\Pi}~,    
\end{equation}
where $\mathbf{P}$ is a matrix containing all the measurement statistics, arising from $D$ input states and yielding $N$ outcomes. The matrix $\mathbf{F}$ contains the photon number distributions of all $D$ probe states, truncated at the maximum Hilbert space dimension $M$. $\mathbf{\Pi}$ is an $M\times N$ matrix, the elements of which are ${\Pi}_{i,n}=\theta_i^{\left(n\right)}$ such that the columns  correspond to the diagonal elements of the POVMs $\pi_n$ given in Eq.~\ref{eqn:pin}, with the sum truncated at $M-1$. Thus, following~\cite{lundeen2009tomography}, detector tomography can be cast as a matrix inversion problem, where the task is to determine the unknown $\mathbf{\Pi}$ from known $\mathbf{P}$ and $\mathbf{F}$.


\subsection{Input state preparation and characterisation}
\begin{figure}[h]
    \centering
    \includegraphics[width=0.45\textwidth]{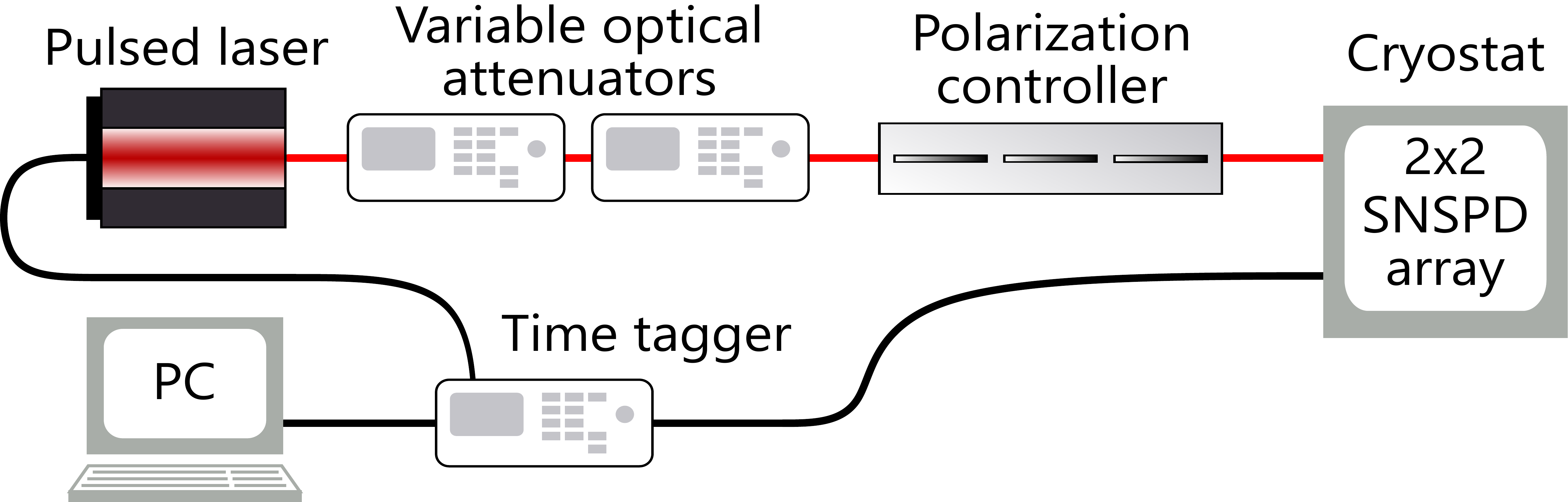}
    \caption{A 1556\,nm pulsed laser produces coherent states at a repetition rate of 500\,kHz, which are then attenuated using computer-controlled variable optical attenuators and detected by a 2$\times$2 array of SNSPDs. A time-tagger 
    is used to measure the electronic response from the detector. For further details see text (Sec.~\ref{sec:definingoutcomes}). 
    }
    \label{fig:sketch}
\end{figure}

The experiment is conducted using the setup shown in Fig.~\ref{fig:sketch}. Experimentally, the set of input states are created from a 1556\,nm laser emitting 9\,ps pulses at a repetition rate of 500\,kHz. The pulse energy is set using two variable optical attenuators. 
Assuming Poissonian statistics of the laser pulses (which was separately determined using an autocorrelation measurement yielding $g^{(2)}(0)=1.00006\left(17\right)$ across the power range), the mean photon number per pulse is determined by 
\begin{equation}
    \bar{n}=-\frac{1}{\eta_{\textrm{cal}}}\ln\left[1-p\left(\textrm{click}\right)\right]~,
\end{equation}
where $\eta_{\textrm{cal}}=83\pm5\,\%$ is the detection efficiency and $p\left(\textrm{click}\right)$ is the click probability, respectively, of the calibration detector (a single-element SNSPD), with the dark-count probability per pulse (of $2\times 10^{-7}$) neglected. A total of $D=19$ coherent states were used, the photon number distributions of which span a range of 0 to 332 photons per pulse. To express these coherent states in a finite matrix $\mathbf{F}$, their dimension was truncated at $M=443$. The elements of $\mathbf{F}$ are given by the Poisson distributions 
\begin{equation}
    \mathrm{F}_{d,i}=\frac{\left|\alpha_d\right|^{2i}}{i!}e^{-\left|\alpha_d\right|^2}~,
\end{equation}
for mean photon numbers $|\alpha_d|^2$, where $d\in\left[0,18\right]$ and $i\in\left[0,442\right]$, such that the maximum dimension was chosen to include probability amplitudes at six standard deviations greater than the largest coherent state
. The magnitudes of the different coherent states are chosen to scale approximately quadratically, {i.e.} $\left|\alpha_d\right|^2\approx d^2$.

\subsection{Defining outcomes} \label{sec:definingoutcomes}
The detector under test is a commercial device (Photon Spot) comprising a 2$\times$2 array of SNSPDs electrically connected in series
. The output of this device is a voltage signal proportional to the number of pixels which fire
. We read out this device using a time-tagger, which counts the number of times a particular voltage threshold is exceeded, with different threshold settings corresponding to the different number of pixels that fire. Further details on the detector itself can be found in Ref.~\cite{tiedau2020single}. 


\begin{figure*}[t!]
    \centering
    \includegraphics[width=1\textwidth]{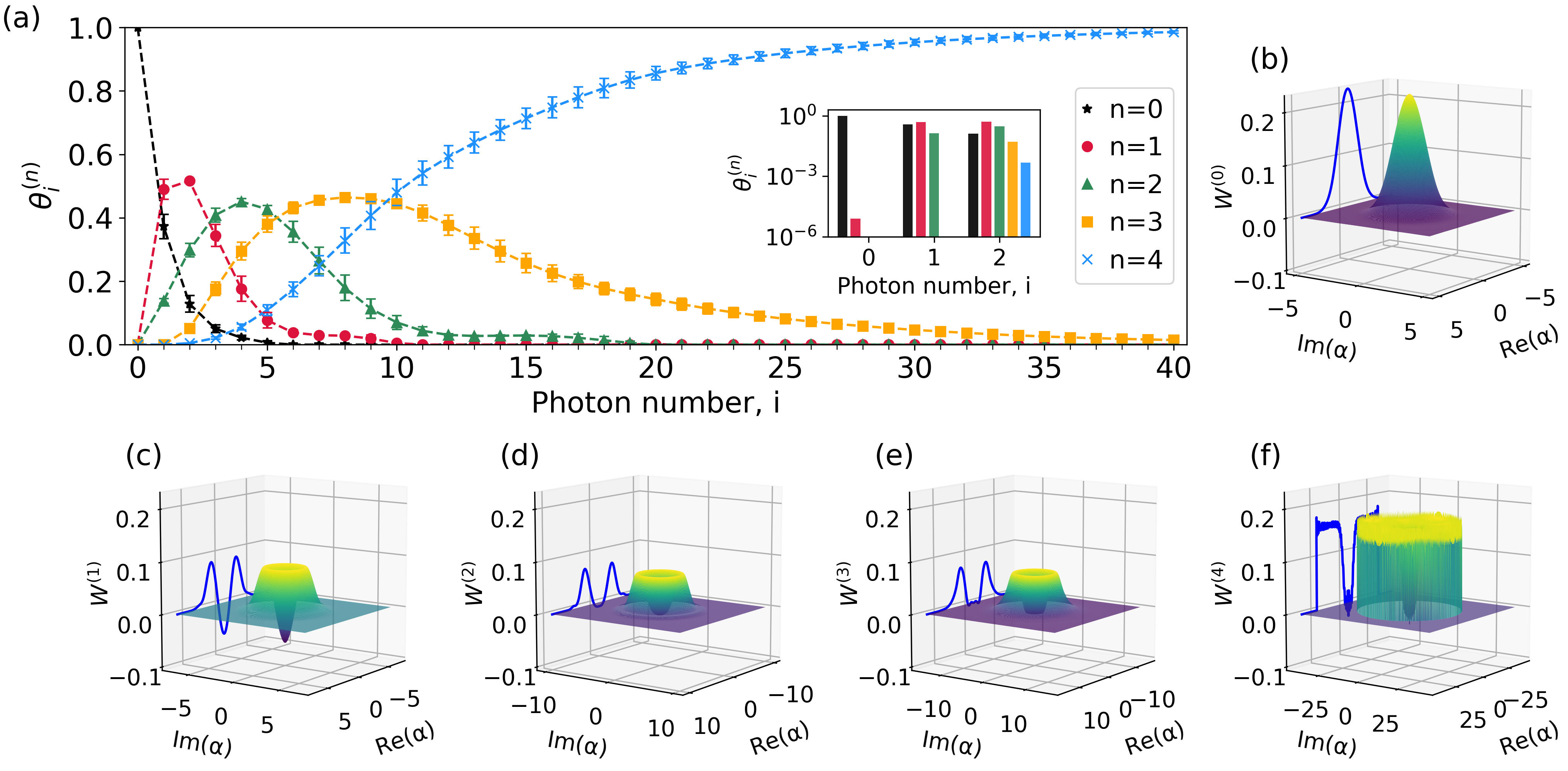}
    \caption{(a) Diagonal elements of the reconstructed POVM operators in the photon-number basis for all five outcomes: vacuum (no-click) (black), one click (red), two clicks (green), three clicks (yellow) and at least four clicks (blue). The inset shows the same data on a log-scale for the first three photon numbers. 
    (b)-(f) Wigner functions of the five possible outcomes of the 2$\times$2 array of SNSPDs, calculated from the POVMs in (a). Clear negativity and the non-Gaussian nature of the one, two and three-click events can be seen.}
    \label{fig:POVMs}
\end{figure*}

We obtain data in an ensemble measurement: we sequentially cycle through the threshold settings, thereby obtaining count rates $c_{n}$, corresponding to at least $n$ pixels firing. For each coherent state amplitude and threshold setting, click statistics were obtained for $5\times10^6$ 
pulses, measured in a coincidence window of 15\,ns synchronised to the  pulse train from the laser. 
In this analysis $c_{0}$ (at least 0 clicks) corresponds to the repetition rate of the experiment. 
Note that these outcomes are not orthogonal, since events contributing to $c_{n}$ are also contained within $c_{j>n}$. Orthogonal outcomes $c_n^\prime$ are obtained 
using the transformation $c^\prime_n=c_n-c_{n+1}$, such that $c_n^\prime$ gives the rate at which exactly $n$ detectors click. The probabilities $P_{d,n}$ of each outcome, evaluated given an input state $d$, are thus given by $P_{d,n}=\frac{c^\prime_n}{c_0}\Big|_d$. The elements $P_{d,n}$ make up the outcome matrix $\mathbf{P}$.

\subsection{Matrix inversion and smoothing}
Given the matrices of input states $\mathbf{F}$ and outcomes $\mathbf{P}$, the matrix $\mathbf{\Pi}$ corresponding to the POVM set $\{\pi_n\}$ can be found by inversion. In order to maintain physical POVMs, this inversion can be recast as the optimisation~\cite{lundeen2009tomography}:
\begin{align}
    \min\left\{\left|\left|\mathbf{P}-\mathbf{F}\mathbf{\Pi}\right|\right|_2+g\left(\mathbf{\Pi}\right)\right\}~,
\end{align}
where $\left|\left|\cdot\right|\right|_2$ indicates the Frobenius norm~\cite{golub2013matrix} and the function
\begin{equation}
g\left(\mathbf{\Pi}\right)=\epsilon \sum_{i,n}\left(\theta^{(n)}_i-\theta^{(n)}_{i+1}\right)^2~,
\end{equation}
scaled by a factor $\epsilon$, ensures that the result is smooth~\cite{feito2009measuring}.  We choose a smoothing parameter of $\epsilon=0.1$; the effects of choosing a different smoothing parameter on the POVM elements are discussed in more detail in Section~\ref{sec:darks}. The code to perform this inversion was written using the CVXPY module in Python~\cite{cvxpy,cvxpy_rewriting} and is available online~\cite{schapeler2020cvxpycode}.

\section{Results}
The reconstructed POVM elements for all five outcomes are shown in Fig.~\ref{fig:POVMs}(a). The inset shows the same data for zero, one and two photons incident, on a log-scale. Assuming phase-insensitive POVMs, Wigner functions corresponding to the five outcomes (0-4 clicks) may also be reconstructed, as shown in Figs.~\ref{fig:POVMs}(b)-\ref{fig:POVMs}(f). Error bars are calculated based on assuming 5\% uncertainty in the amplitudes of the coherent states. This reflects the uncertainty in the calibration procedure; uncertainty due to finite counting statistics is negligible in comparison.

Since the elements $\theta^{(n)}_i$ are the conditional probabilities 
\begin{equation}
    \theta^{(n)}_i=p\left(n~\textrm{clicks}|i~\textrm{incident photons}\right)~,
\end{equation}
the reconstruction can yield bounds on detector parameters such as dark counts, cross-talk and overall efficiency. Crucially, this can be achieved without any underlying assumptions about the detector geometry or circuitry.

\subsection{Efficiency}
Efficiency is intuitively defined as the probability that the detector clicks given that a single photon was incident, {i.e.} 
\begin{equation}
\eta=\sum_{n=1}^{N-1}p\left(n|1\right)=1-p\left(0|1\right)
\end{equation}
where $N$ is the total number of outcomes. For our detector, $p\left(0|1\right)=0.37\pm0.04$, which  results in an efficiency of $\eta=63\%\pm4\%$. The error arises from the uncertainty in the determining the mean photon number used for each of the input states. This agrees well with the independently measured efficiency of $\eta=65\%\pm4\%$, based on direct comparison with a calibrated single-pixel detector for fixed incident power. The uncertainty in this measurement stems from the calibration procedure. 

\subsection{Dark counts}\label{sec:darks}
As shown in the inset in Fig.~\ref{fig:POVMs}(a), the dark-count probability is defined as the single-click probability (red) when zero photons are incident. 
For the device as a whole, this corresponds to the probability
\begin{equation}
    p_\textrm{dark}=p\left(1|0\right)~.
\end{equation}
For the detector under test, the element $p\left(1|0\right)=\left(5.9\pm1.6\right)\times10^{-6}$. This agrees well given the dark count probability measured independently to be $p_\textrm{dark}=\left(6.34\pm0.15\right)\times10^{-6}$. The close agreement clearly shows that the intuitive definition of how dark counts manifest in the POVM elements is reasonable.


The importance of choosing an appropriate smoothing parameter is manifest in the dark count estimation. 
The smoothing factor $\epsilon$ can take values between zero and one. Previous work~\cite{feito2009measuring} has shown that the value itself is relatively unimportant, since the error associated with the reconstruction is largely independent of the smoothing factor. However, in cases where neighboring POVM elements are expected to vary by several orders of magnitude, 
choosing a smoothing factor that is too large may significantly overestimate the smaller of the two elements. SNSPDs are a pertinent example of a detector whose tomographic reconstruction may be susceptible to an inopportune choice of the smoothing factor. 

\begin{figure}[h]
    \centering
    \includegraphics[width=0.45\textwidth]{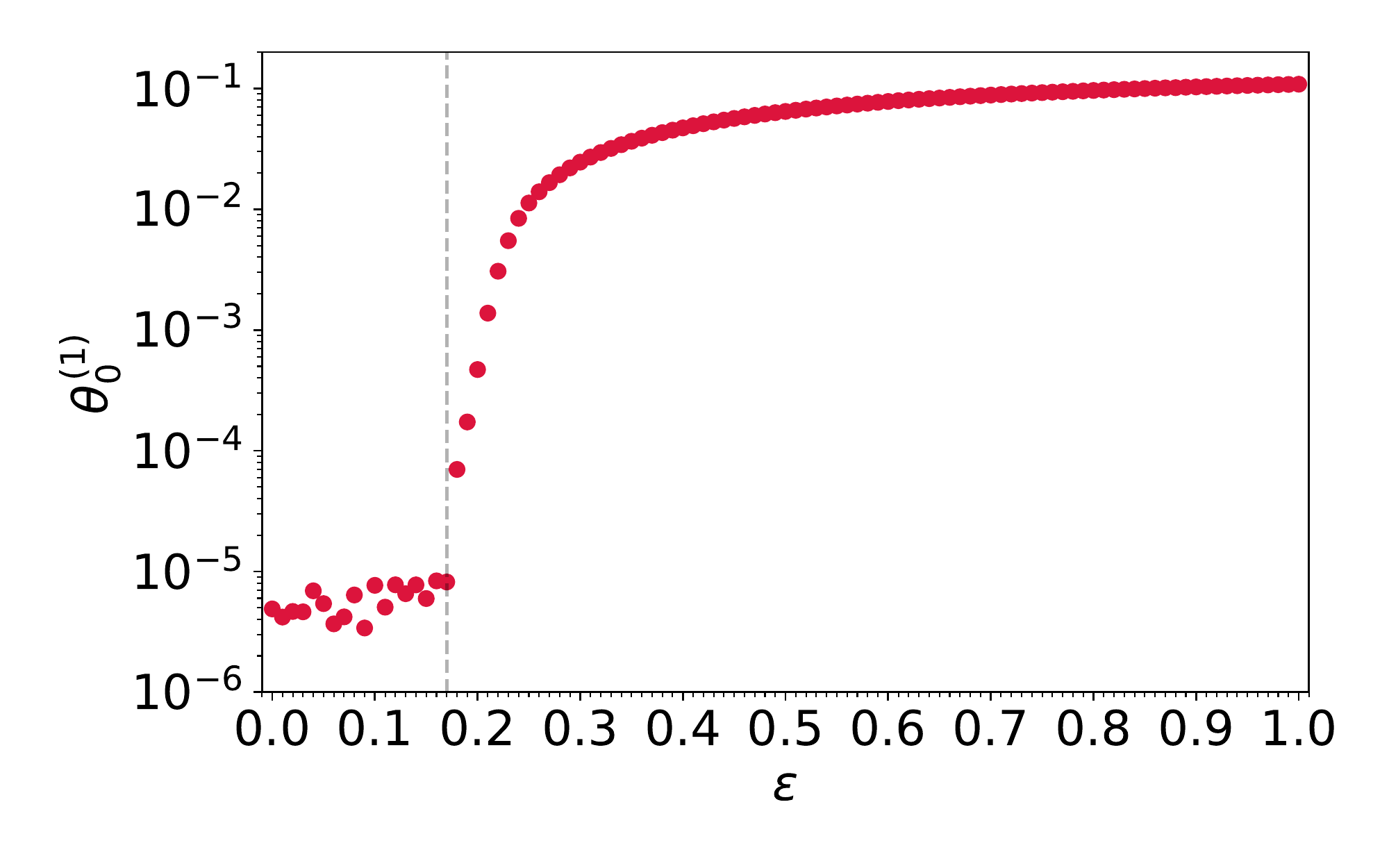}
    \caption{Dependence of the $\theta_0^{(1)}$ POVM element on the smoothing factor $\epsilon$. Above a factor of $\epsilon=0.17$ (indicated by the dashed line) the smoothing causes an overestimation of the given element.}
    \label{fig:smoothing}
\end{figure}

To illustrate this, in Fig.~\ref{fig:smoothing} we plot the element $\theta_0^{(1)}$ as a function of smoothing factor $\epsilon$. Below a threshold of $\epsilon=0.17$, the POVM element is independent of $\epsilon$; however, above this point, the smoothing factor causes an overestimate. We arbitrarily choose a smoothing factor of $\epsilon=0.1$ to be below this threshold. 

\subsection{Cross-talk}
In contrast to dark counts, cross-talk is regarded as conditional noise: additional counts arising due to a pixel firing. As can be seen in the inset to Fig.~\ref{fig:POVMs}(a), cross-talk is most clearly manifest in the much larger probability of two clicks (green) given one photon (compared with one click given zero photons). In principle, these effects contributes to all POVM elements $p\left(j|i\right)$, for $j>\textrm{max}\left(i,1\right)$. For example, one POVM element containing this effect is the element $p\left(2|1\right)$, {i.e.} where one incident photon causes two clicks. 
In the absence of cross-talk, one expects the $p\left(2|1\right)$ term to comprise only clicks given a photon is incident $p\left(1|1\right)$, and a single dark count arising from the three remaining detectors
. Any additional counts in the scenario we attribute to cross-talk. As such, we define the single-pixel cross-talk probability as
\begin{equation}
    p_\textrm{xtalk}=p\left(2|1\right)-p\left(1|1\right)p\left(1|0\right)~.
\end{equation}
In our case, $p\left(2|1\right)=0.14\pm0.01$, $p\left(1|1\right)=0.49\pm0.03$ and $p\left(1|0\right)=\left(5.9\pm1.6\right)\times10^{-6}$ from above. We therefore estimate a single-pixel cross-talk probability of $14\pm1$\%, which is dominated by the leading term since the dark-count probability is significantly smaller. 


In general, an independent measurement and/or modelling of the cross-talk probability is very challenging, since it depends on a range of factors, including the number of pixels which have fired, the number and location of remaining pixels, and correlations between particular pixels~\cite{eraerds2007sipm,akiba2008multipixel,afek2009quantum,ramilli2010photon,kalashnikov2011accessing,dovrat2012measurements,kalashnikov2012crosstalk,nehra2020photon}. Nevertheless, our method can provide an estimate of this probability in a relatively straightforward manner.

\section{Conclusion}
Quantum detector tomography is a powerful tool to characterise a measurement process without recourse to the underlying physics of the detector. Nevertheless, certain physical properties of a detector can be inferred, such as efficiency, dark counts, and cross talk, where the latter is exclusive to array detectors. We applied this technique to characterise a commercial four-pixel array of superconducting nanowire single photon detectors, and were able to identify these figures of merit directly using just four POVM elements. This is particularly useful for identifying cross-talk probability, which may be otherwise difficult to do based on underlying models of the detector electronics.
Furthermore, as the size of arrays become increasingly large, the need to characterise the device as a whole, rather than on a per-pixel basis, will become increasingly important.

\section*{Acknowledgements}
We are grateful to Vikas Anant of Photon Spot, Inc. for the detector, and to Thomas Hummel for discussions.

\section*{Funding}
This project is supported by the German Federal Ministry of Education and Research (BMBF) under the funding program Photonics Research Germany, grant number 13N14911.

\section*{Disclosures}
The authors declare no conflicts of interest.


\bibliography{sources.bib}

\begin{thebibliography}{51}%
\makeatletter
\providecommand \@ifxundefined [1]{%
 \@ifx{#1\undefined}
}%
\providecommand \@ifnum [1]{%
 \ifnum #1\expandafter \@firstoftwo
 \else \expandafter \@secondoftwo
 \fi
}%
\providecommand \@ifx [1]{%
 \ifx #1\expandafter \@firstoftwo
 \else \expandafter \@secondoftwo
 \fi
}%
\providecommand \natexlab [1]{#1}%
\providecommand \enquote  [1]{``#1''}%
\providecommand \bibnamefont  [1]{#1}%
\providecommand \bibfnamefont [1]{#1}%
\providecommand \citenamefont [1]{#1}%
\providecommand \href@noop [0]{\@secondoftwo}%
\providecommand \href [0]{\begingroup \@sanitize@url \@href}%
\providecommand \@href[1]{\@@startlink{#1}\@@href}%
\providecommand \@@href[1]{\endgroup#1\@@endlink}%
\providecommand \@sanitize@url [0]{\catcode `\\12\catcode `\$12\catcode
  `\&12\catcode `\#12\catcode `\^12\catcode `\_12\catcode `\%12\relax}%
\providecommand \@@startlink[1]{}%
\providecommand \@@endlink[0]{}%
\providecommand \url  [0]{\begingroup\@sanitize@url \@url }%
\providecommand \@url [1]{\endgroup\@href {#1}{\urlprefix }}%
\providecommand \urlprefix  [0]{URL }%
\providecommand \Eprint [0]{\href }%
\providecommand \doibase [0]{https://doi.org/}%
\providecommand \selectlanguage [0]{\@gobble}%
\providecommand \bibinfo  [0]{\@secondoftwo}%
\providecommand \bibfield  [0]{\@secondoftwo}%
\providecommand \translation [1]{[#1]}%
\providecommand \BibitemOpen [0]{}%
\providecommand \bibitemStop [0]{}%
\providecommand \bibitemNoStop [0]{.\EOS\space}%
\providecommand \EOS [0]{\spacefactor3000\relax}%
\providecommand \BibitemShut  [1]{\csname bibitem#1\endcsname}%
\let\auto@bib@innerbib\@empty
\bibitem [{\citenamefont {Gol'Tsman}\ \emph {et~al.}(2001)\citenamefont
  {Gol'Tsman}, \citenamefont {Okunev}, \citenamefont {Chulkova}, \citenamefont
  {Lipatov}, \citenamefont {Semenov}, \citenamefont {Smirnov}, \citenamefont
  {Voronov}, \citenamefont {Dzardanov}, \citenamefont {Williams},\ and\
  \citenamefont {Sobolewski}}]{gol2001picosecond}%
  \BibitemOpen
  \bibfield  {author} {\bibinfo {author} {\bibfnamefont {G.}~\bibnamefont
  {Gol'Tsman}}, \bibinfo {author} {\bibfnamefont {O.}~\bibnamefont {Okunev}},
  \bibinfo {author} {\bibfnamefont {G.}~\bibnamefont {Chulkova}}, \bibinfo
  {author} {\bibfnamefont {A.}~\bibnamefont {Lipatov}}, \bibinfo {author}
  {\bibfnamefont {A.}~\bibnamefont {Semenov}}, \bibinfo {author} {\bibfnamefont
  {K.}~\bibnamefont {Smirnov}}, \bibinfo {author} {\bibfnamefont
  {B.}~\bibnamefont {Voronov}}, \bibinfo {author} {\bibfnamefont
  {A.}~\bibnamefont {Dzardanov}}, \bibinfo {author} {\bibfnamefont
  {C.}~\bibnamefont {Williams}},\ and\ \bibinfo {author} {\bibfnamefont
  {R.}~\bibnamefont {Sobolewski}},\ }\bibfield  {title} {\bibinfo {title}
  {Picosecond superconducting single-photon optical detector},\ }\href@noop {}
  {\bibfield  {journal} {\bibinfo  {journal} {Applied Physics Letters}\
  }\textbf {\bibinfo {volume} {79}},\ \bibinfo {pages} {705} (\bibinfo {year}
  {2001})}\BibitemShut {NoStop}%
\bibitem [{\citenamefont {Natarajan}\ \emph {et~al.}(2012)\citenamefont
  {Natarajan}, \citenamefont {Tanner},\ and\ \citenamefont
  {Hadfield}}]{natarajan_superconducting_2012}%
  \BibitemOpen
  \bibfield  {author} {\bibinfo {author} {\bibfnamefont {C.~M.}\ \bibnamefont
  {Natarajan}}, \bibinfo {author} {\bibfnamefont {M.~G.}\ \bibnamefont
  {Tanner}},\ and\ \bibinfo {author} {\bibfnamefont {R.~H.}\ \bibnamefont
  {Hadfield}},\ }\bibfield  {title} {{\selectlanguage {english}\bibinfo {title}
  {Superconducting nanowire single-photon detectors: physics and
  applications}},\ }\href {https://doi.org/10.1088/0953-2048/25/6/063001}
  {\bibfield  {journal} {\bibinfo  {journal} {Superconductor Science and
  Technology}\ }\textbf {\bibinfo {volume} {25}},\ \bibinfo {pages} {063001}
  (\bibinfo {year} {2012})}\BibitemShut {NoStop}%
\bibitem [{\citenamefont {Marsili}\ \emph {et~al.}(2013)\citenamefont
  {Marsili}, \citenamefont {Verma}, \citenamefont {Stern}, \citenamefont
  {Harrington}, \citenamefont {Lita}, \citenamefont {Gerrits}, \citenamefont
  {Vayshenker}, \citenamefont {Baek}, \citenamefont {Shaw}, \citenamefont
  {Mirin},\ and\ \citenamefont {Nam}}]{marsili_detecting_2013}%
  \BibitemOpen
  \bibfield  {author} {\bibinfo {author} {\bibfnamefont {F.}~\bibnamefont
  {Marsili}}, \bibinfo {author} {\bibfnamefont {V.~B.}\ \bibnamefont {Verma}},
  \bibinfo {author} {\bibfnamefont {J.~A.}\ \bibnamefont {Stern}}, \bibinfo
  {author} {\bibfnamefont {S.}~\bibnamefont {Harrington}}, \bibinfo {author}
  {\bibfnamefont {A.~E.}\ \bibnamefont {Lita}}, \bibinfo {author}
  {\bibfnamefont {T.}~\bibnamefont {Gerrits}}, \bibinfo {author} {\bibfnamefont
  {I.}~\bibnamefont {Vayshenker}}, \bibinfo {author} {\bibfnamefont
  {B.}~\bibnamefont {Baek}}, \bibinfo {author} {\bibfnamefont {M.~D.}\
  \bibnamefont {Shaw}}, \bibinfo {author} {\bibfnamefont {R.~P.}\ \bibnamefont
  {Mirin}},\ and\ \bibinfo {author} {\bibfnamefont {S.~W.}\ \bibnamefont
  {Nam}},\ }\bibfield  {title} {{\selectlanguage {english}\bibinfo {title}
  {Detecting single infrared photons with 93\% system efficiency}},\ }\href
  {https://doi.org/10.1038/nphoton.2013.13} {\bibfield  {journal} {\bibinfo
  {journal} {Nature Photonics}\ }\textbf {\bibinfo {volume} {7}},\ \bibinfo
  {pages} {210} (\bibinfo {year} {2013})}\BibitemShut {NoStop}%
\bibitem [{\citenamefont {Esmaeil~Zadeh}\ \emph {et~al.}(2017)\citenamefont
  {Esmaeil~Zadeh}, \citenamefont {Los}, \citenamefont {Gourgues}, \citenamefont
  {Steinmetz}, \citenamefont {Bulgarini}, \citenamefont {Dobrovolskiy},
  \citenamefont {Zwiller},\ and\ \citenamefont
  {Dorenbos}}]{esmaeil_zadeh_single-photon_2017}%
  \BibitemOpen
  \bibfield  {author} {\bibinfo {author} {\bibfnamefont {I.}~\bibnamefont
  {Esmaeil~Zadeh}}, \bibinfo {author} {\bibfnamefont {J.~W.~N.}\ \bibnamefont
  {Los}}, \bibinfo {author} {\bibfnamefont {R.~B.~M.}\ \bibnamefont
  {Gourgues}}, \bibinfo {author} {\bibfnamefont {V.}~\bibnamefont {Steinmetz}},
  \bibinfo {author} {\bibfnamefont {G.}~\bibnamefont {Bulgarini}}, \bibinfo
  {author} {\bibfnamefont {S.~M.}\ \bibnamefont {Dobrovolskiy}}, \bibinfo
  {author} {\bibfnamefont {V.}~\bibnamefont {Zwiller}},\ and\ \bibinfo {author}
  {\bibfnamefont {S.~N.}\ \bibnamefont {Dorenbos}},\ }\bibfield  {title}
  {\bibinfo {title} {Single-photon detectors combining high efficiency, high
  detection rates, and ultra-high timing resolution},\ }\href
  {https://doi.org/10.1063/1.5000001} {\bibfield  {journal} {\bibinfo
  {journal} {APL Photonics}\ }\textbf {\bibinfo {volume} {2}},\ \bibinfo
  {pages} {111301} (\bibinfo {year} {2017})}\BibitemShut {NoStop}%
\bibitem [{\citenamefont {Korzh}\ \emph {et~al.}(2020)\citenamefont {Korzh},
  \citenamefont {Zhao}, \citenamefont {Allmaras}, \citenamefont {Frasca},
  \citenamefont {Autry}, \citenamefont {Bersin}, \citenamefont {Beyer},
  \citenamefont {Briggs}, \citenamefont {Bumble}, \citenamefont {Colangelo}
  \emph {et~al.}}]{korzh2020demonstration}%
  \BibitemOpen
  \bibfield  {author} {\bibinfo {author} {\bibfnamefont {B.}~\bibnamefont
  {Korzh}}, \bibinfo {author} {\bibfnamefont {Q.-Y.}\ \bibnamefont {Zhao}},
  \bibinfo {author} {\bibfnamefont {J.~P.}\ \bibnamefont {Allmaras}}, \bibinfo
  {author} {\bibfnamefont {S.}~\bibnamefont {Frasca}}, \bibinfo {author}
  {\bibfnamefont {T.~M.}\ \bibnamefont {Autry}}, \bibinfo {author}
  {\bibfnamefont {E.~A.}\ \bibnamefont {Bersin}}, \bibinfo {author}
  {\bibfnamefont {A.~D.}\ \bibnamefont {Beyer}}, \bibinfo {author}
  {\bibfnamefont {R.~M.}\ \bibnamefont {Briggs}}, \bibinfo {author}
  {\bibfnamefont {B.}~\bibnamefont {Bumble}}, \bibinfo {author} {\bibfnamefont
  {M.}~\bibnamefont {Colangelo}}, \emph {et~al.},\ }\bibfield  {title}
  {\bibinfo {title} {Demonstration of sub-3 ps temporal resolution with a
  superconducting nanowire single-photon detector},\ }\href@noop {} {\bibfield
  {journal} {\bibinfo  {journal} {Nature Photonics}\ }\textbf {\bibinfo
  {volume} {14}},\ \bibinfo {pages} {250} (\bibinfo {year} {2020})}\BibitemShut
  {NoStop}%
\bibitem [{\citenamefont {McCaughan}(2018)}]{mccaughan_readout_2018}%
  \BibitemOpen
  \bibfield  {author} {\bibinfo {author} {\bibfnamefont {A.~N.}\ \bibnamefont
  {McCaughan}},\ }\bibfield  {title} {\bibinfo {title} {Readout architectures
  for superconducting nanowire single photon detectors},\ }\bibfield  {journal}
  {\bibinfo  {journal} {Supercond. Sci. Technol.}\ }\textbf {\bibinfo {volume}
  {31}},\ \href {https://doi.org/10.1088/1361-6668/aaa1b3}
  {10.1088/1361-6668/aaa1b3} (\bibinfo {year} {2018})\BibitemShut {NoStop}%
\bibitem [{\citenamefont {Miyajima}\ \emph {et~al.}(2018)\citenamefont
  {Miyajima}, \citenamefont {Yabuno}, \citenamefont {Miki}, \citenamefont
  {Yamashita},\ and\ \citenamefont {Terai}}]{miyajima2018high}%
  \BibitemOpen
  \bibfield  {author} {\bibinfo {author} {\bibfnamefont {S.}~\bibnamefont
  {Miyajima}}, \bibinfo {author} {\bibfnamefont {M.}~\bibnamefont {Yabuno}},
  \bibinfo {author} {\bibfnamefont {S.}~\bibnamefont {Miki}}, \bibinfo {author}
  {\bibfnamefont {T.}~\bibnamefont {Yamashita}},\ and\ \bibinfo {author}
  {\bibfnamefont {H.}~\bibnamefont {Terai}},\ }\bibfield  {title} {\bibinfo
  {title} {High-time-resolved 64-channel single-flux quantum-based address
  encoder integrated with a multi-pixel superconducting nanowire single-photon
  detector},\ }\href@noop {} {\bibfield  {journal} {\bibinfo  {journal} {Opt.
  Express}\ }\textbf {\bibinfo {volume} {26}},\ \bibinfo {pages} {29045}
  (\bibinfo {year} {2018})}\BibitemShut {NoStop}%
\bibitem [{\citenamefont {Cahall}\ \emph {et~al.}(2018)\citenamefont {Cahall},
  \citenamefont {Gauthier},\ and\ \citenamefont {Kim}}]{cahall2018scalable}%
  \BibitemOpen
  \bibfield  {author} {\bibinfo {author} {\bibfnamefont {C.}~\bibnamefont
  {Cahall}}, \bibinfo {author} {\bibfnamefont {D.~J.}\ \bibnamefont
  {Gauthier}},\ and\ \bibinfo {author} {\bibfnamefont {J.}~\bibnamefont
  {Kim}},\ }\bibfield  {title} {\bibinfo {title} {Scalable cryogenic readout
  circuit for a superconducting nanowire single-photon detector system},\
  }\href@noop {} {\bibfield  {journal} {\bibinfo  {journal} {Rev. Sci.
  Instrum.}\ }\textbf {\bibinfo {volume} {89}},\ \bibinfo {pages} {063117}
  (\bibinfo {year} {2018})}\BibitemShut {NoStop}%
\bibitem [{\citenamefont {Gaggero}\ \emph {et~al.}(2019)\citenamefont
  {Gaggero}, \citenamefont {Martini}, \citenamefont {Mattioli}, \citenamefont
  {Chiarello}, \citenamefont {Cernansky}, \citenamefont {Politi},\ and\
  \citenamefont {Leoni}}]{gaggero2019amplitude}%
  \BibitemOpen
  \bibfield  {author} {\bibinfo {author} {\bibfnamefont {A.}~\bibnamefont
  {Gaggero}}, \bibinfo {author} {\bibfnamefont {F.}~\bibnamefont {Martini}},
  \bibinfo {author} {\bibfnamefont {F.}~\bibnamefont {Mattioli}}, \bibinfo
  {author} {\bibfnamefont {F.}~\bibnamefont {Chiarello}}, \bibinfo {author}
  {\bibfnamefont {R.}~\bibnamefont {Cernansky}}, \bibinfo {author}
  {\bibfnamefont {A.}~\bibnamefont {Politi}},\ and\ \bibinfo {author}
  {\bibfnamefont {R.}~\bibnamefont {Leoni}},\ }\bibfield  {title} {\bibinfo
  {title} {Amplitude-multiplexed readout of single photon detectors based on
  superconducting nanowires},\ }\href@noop {} {\bibfield  {journal} {\bibinfo
  {journal} {Optica}\ }\textbf {\bibinfo {volume} {6}},\ \bibinfo {pages} {823}
  (\bibinfo {year} {2019})}\BibitemShut {NoStop}%
\bibitem [{\citenamefont {Tiedau}\ \emph {et~al.}(2020)\citenamefont {Tiedau},
  \citenamefont {Schapeler}, \citenamefont {Anant}, \citenamefont {Fedder},
  \citenamefont {Silberhorn},\ and\ \citenamefont
  {Bartley}}]{tiedau2020single}%
  \BibitemOpen
  \bibfield  {author} {\bibinfo {author} {\bibfnamefont {J.}~\bibnamefont
  {Tiedau}}, \bibinfo {author} {\bibfnamefont {T.}~\bibnamefont {Schapeler}},
  \bibinfo {author} {\bibfnamefont {V.}~\bibnamefont {Anant}}, \bibinfo
  {author} {\bibfnamefont {H.}~\bibnamefont {Fedder}}, \bibinfo {author}
  {\bibfnamefont {C.}~\bibnamefont {Silberhorn}},\ and\ \bibinfo {author}
  {\bibfnamefont {T.~J.}\ \bibnamefont {Bartley}},\ }\bibfield  {title}
  {\bibinfo {title} {Single-channel electronic readout of a multipixel
  superconducting nanowire single photon detector},\ }\href@noop {} {\bibfield
  {journal} {\bibinfo  {journal} {Optics Express}\ }\textbf {\bibinfo {volume}
  {28}},\ \bibinfo {pages} {5528} (\bibinfo {year} {2020})}\BibitemShut
  {NoStop}%
\bibitem [{\citenamefont {Allmaras}\ \emph {et~al.}(2020)\citenamefont
  {Allmaras}, \citenamefont {Wollman}, \citenamefont {Beyer}, \citenamefont
  {Briggs}, \citenamefont {Korzh}, \citenamefont {Bumble},\ and\ \citenamefont
  {Shaw}}]{allmaras2020demonstration}%
  \BibitemOpen
  \bibfield  {author} {\bibinfo {author} {\bibfnamefont {J.~P.}\ \bibnamefont
  {Allmaras}}, \bibinfo {author} {\bibfnamefont {E.~E.}\ \bibnamefont
  {Wollman}}, \bibinfo {author} {\bibfnamefont {A.~D.}\ \bibnamefont {Beyer}},
  \bibinfo {author} {\bibfnamefont {R.~M.}\ \bibnamefont {Briggs}}, \bibinfo
  {author} {\bibfnamefont {B.~A.}\ \bibnamefont {Korzh}}, \bibinfo {author}
  {\bibfnamefont {B.}~\bibnamefont {Bumble}},\ and\ \bibinfo {author}
  {\bibfnamefont {M.~D.}\ \bibnamefont {Shaw}},\ }\bibfield  {title} {\bibinfo
  {title} {Demonstration of a thermally coupled row-column snspd imaging
  array},\ }\href@noop {} {\bibfield  {journal} {\bibinfo  {journal} {Nano
  Lett.}\ }\textbf {\bibinfo {volume} {20}},\ \bibinfo {pages} {2163} (\bibinfo
  {year} {2020})}\BibitemShut {NoStop}%
\bibitem [{\citenamefont {Takeuchi}\ \emph {et~al.}(2020)\citenamefont
  {Takeuchi}, \citenamefont {China}, \citenamefont {Miki}, \citenamefont
  {Miyajima}, \citenamefont {Yabuno}, \citenamefont {Yoshikawa},\ and\
  \citenamefont {Terai}}]{takeuchi2020scalable}%
  \BibitemOpen
  \bibfield  {author} {\bibinfo {author} {\bibfnamefont {N.}~\bibnamefont
  {Takeuchi}}, \bibinfo {author} {\bibfnamefont {F.}~\bibnamefont {China}},
  \bibinfo {author} {\bibfnamefont {S.}~\bibnamefont {Miki}}, \bibinfo {author}
  {\bibfnamefont {S.}~\bibnamefont {Miyajima}}, \bibinfo {author}
  {\bibfnamefont {M.}~\bibnamefont {Yabuno}}, \bibinfo {author} {\bibfnamefont
  {N.}~\bibnamefont {Yoshikawa}},\ and\ \bibinfo {author} {\bibfnamefont
  {H.}~\bibnamefont {Terai}},\ }\bibfield  {title} {\bibinfo {title} {Scalable
  readout interface for superconducting nanowire single-photon detectors using
  aqfp and rsfq logic families},\ }\href@noop {} {\bibfield  {journal}
  {\bibinfo  {journal} {Opt. Express}\ }\textbf {\bibinfo {volume} {28}},\
  \bibinfo {pages} {15824} (\bibinfo {year} {2020})}\BibitemShut {NoStop}%
\bibitem [{\citenamefont {Dauler}\ \emph {et~al.}(2007)\citenamefont {Dauler},
  \citenamefont {Robinson}, \citenamefont {Kerman}, \citenamefont {Yang},
  \citenamefont {Rosfjord}, \citenamefont {Anant}, \citenamefont {Voronov},
  \citenamefont {Gol'tsman},\ and\ \citenamefont
  {Berggren}}]{dauler_multi-element_2007}%
  \BibitemOpen
  \bibfield  {author} {\bibinfo {author} {\bibfnamefont {E.~A.}\ \bibnamefont
  {Dauler}}, \bibinfo {author} {\bibfnamefont {B.~S.}\ \bibnamefont
  {Robinson}}, \bibinfo {author} {\bibfnamefont {A.~J.}\ \bibnamefont
  {Kerman}}, \bibinfo {author} {\bibfnamefont {J.~K.~W.}\ \bibnamefont {Yang}},
  \bibinfo {author} {\bibfnamefont {K.~M.}\ \bibnamefont {Rosfjord}}, \bibinfo
  {author} {\bibfnamefont {V.}~\bibnamefont {Anant}}, \bibinfo {author}
  {\bibfnamefont {B.}~\bibnamefont {Voronov}}, \bibinfo {author} {\bibfnamefont
  {G.}~\bibnamefont {Gol'tsman}},\ and\ \bibinfo {author} {\bibfnamefont
  {K.~K.}\ \bibnamefont {Berggren}},\ }\bibfield  {title} {\bibinfo {title}
  {Multi-{Element} {Superconducting} {Nanowire} {Single}-{Photon} {Detector}},\
  }\href {https://doi.org/10.1109/TASC.2007.897372} {\bibfield  {journal}
  {\bibinfo  {journal} {IEEE Trans. Appl. Supercond.}\ }\textbf {\bibinfo
  {volume} {17}},\ \bibinfo {pages} {279} (\bibinfo {year} {2007})}\BibitemShut
  {NoStop}%
\bibitem [{\citenamefont {Divochiy}\ \emph {et~al.}(2008)\citenamefont
  {Divochiy}, \citenamefont {Marsili}, \citenamefont {Bitauld}, \citenamefont
  {Gaggero}, \citenamefont {Leoni}, \citenamefont {Mattioli}, \citenamefont
  {Korneev}, \citenamefont {Seleznev}, \citenamefont {Kaurova}, \citenamefont
  {Minaeva}, \citenamefont {Gol'tsman}, \citenamefont {Lagoudakis},
  \citenamefont {Benkhaoul}, \citenamefont {L\'{e}vy},\ and\ \citenamefont
  {Fiore}}]{divochiy_superconducting_2008}%
  \BibitemOpen
  \bibfield  {author} {\bibinfo {author} {\bibfnamefont {A.}~\bibnamefont
  {Divochiy}}, \bibinfo {author} {\bibfnamefont {F.}~\bibnamefont {Marsili}},
  \bibinfo {author} {\bibfnamefont {D.}~\bibnamefont {Bitauld}}, \bibinfo
  {author} {\bibfnamefont {A.}~\bibnamefont {Gaggero}}, \bibinfo {author}
  {\bibfnamefont {R.}~\bibnamefont {Leoni}}, \bibinfo {author} {\bibfnamefont
  {F.}~\bibnamefont {Mattioli}}, \bibinfo {author} {\bibfnamefont
  {A.}~\bibnamefont {Korneev}}, \bibinfo {author} {\bibfnamefont
  {V.}~\bibnamefont {Seleznev}}, \bibinfo {author} {\bibfnamefont
  {N.}~\bibnamefont {Kaurova}}, \bibinfo {author} {\bibfnamefont
  {O.}~\bibnamefont {Minaeva}}, \bibinfo {author} {\bibfnamefont
  {G.}~\bibnamefont {Gol'tsman}}, \bibinfo {author} {\bibfnamefont {K.~G.}\
  \bibnamefont {Lagoudakis}}, \bibinfo {author} {\bibfnamefont
  {M.}~\bibnamefont {Benkhaoul}}, \bibinfo {author} {\bibfnamefont
  {F.}~\bibnamefont {L\'{e}vy}},\ and\ \bibinfo {author} {\bibfnamefont
  {A.}~\bibnamefont {Fiore}},\ }\bibfield  {title} {{\selectlanguage
  {english}\bibinfo {title} {Superconducting nanowire photon-number-resolving
  detector at telecommunication wavelengths}},\ }\href
  {https://doi.org/10.1038/nphoton.2008.51} {\bibfield  {journal} {\bibinfo
  {journal} {Nature Photonics}\ }\textbf {\bibinfo {volume} {2}},\ \bibinfo
  {pages} {302} (\bibinfo {year} {2008})}\BibitemShut {NoStop}%
\bibitem [{\citenamefont {Marsili}\ \emph {et~al.}(2009)\citenamefont
  {Marsili}, \citenamefont {Bitauld}, \citenamefont {Gaggero}, \citenamefont
  {Jahanmirinejad}, \citenamefont {Leoni}, \citenamefont {Mattioli},\ and\
  \citenamefont {Fiore}}]{marsili_physics_2009}%
  \BibitemOpen
  \bibfield  {author} {\bibinfo {author} {\bibfnamefont {F.}~\bibnamefont
  {Marsili}}, \bibinfo {author} {\bibfnamefont {D.}~\bibnamefont {Bitauld}},
  \bibinfo {author} {\bibfnamefont {A.}~\bibnamefont {Gaggero}}, \bibinfo
  {author} {\bibfnamefont {S.}~\bibnamefont {Jahanmirinejad}}, \bibinfo
  {author} {\bibfnamefont {R.}~\bibnamefont {Leoni}}, \bibinfo {author}
  {\bibfnamefont {F.}~\bibnamefont {Mattioli}},\ and\ \bibinfo {author}
  {\bibfnamefont {A.}~\bibnamefont {Fiore}},\ }\bibfield  {title}
  {{\selectlanguage {english}\bibinfo {title} {Physics and application of
  photon number resolving detectors based on superconducting parallel
  nanowires}},\ }\href {https://doi.org/10.1088/1367-2630/11/4/045022}
  {\bibfield  {journal} {\bibinfo  {journal} {New Journal of Physics}\ }\textbf
  {\bibinfo {volume} {11}},\ \bibinfo {pages} {045022} (\bibinfo {year}
  {2009})}\BibitemShut {NoStop}%
\bibitem [{\citenamefont {Jahanmirinejad}\ \emph {et~al.}(2012)\citenamefont
  {Jahanmirinejad}, \citenamefont {Frucci}, \citenamefont {Mattioli},
  \citenamefont {Sahin}, \citenamefont {Gaggero}, \citenamefont {Leoni},\ and\
  \citenamefont {Fiore}}]{jahanmirinejad_photon-number_2012}%
  \BibitemOpen
  \bibfield  {author} {\bibinfo {author} {\bibfnamefont {S.}~\bibnamefont
  {Jahanmirinejad}}, \bibinfo {author} {\bibfnamefont {G.}~\bibnamefont
  {Frucci}}, \bibinfo {author} {\bibfnamefont {F.}~\bibnamefont {Mattioli}},
  \bibinfo {author} {\bibfnamefont {D.}~\bibnamefont {Sahin}}, \bibinfo
  {author} {\bibfnamefont {A.}~\bibnamefont {Gaggero}}, \bibinfo {author}
  {\bibfnamefont {R.}~\bibnamefont {Leoni}},\ and\ \bibinfo {author}
  {\bibfnamefont {A.}~\bibnamefont {Fiore}},\ }\bibfield  {title} {\bibinfo
  {title} {Photon-number resolving detector based on a series array of
  superconducting nanowires},\ }\href {https://doi.org/10.1063/1.4746248}
  {\bibfield  {journal} {\bibinfo  {journal} {Applied Physics Letters}\
  }\textbf {\bibinfo {volume} {101}},\ \bibinfo {pages} {072602} (\bibinfo
  {year} {2012})}\BibitemShut {NoStop}%
\bibitem [{\citenamefont {Zhao}\ \emph {et~al.}(2013)\citenamefont {Zhao},
  \citenamefont {McCaughan}, \citenamefont {Bellei}, \citenamefont {Najafi},
  \citenamefont {De~Fazio}, \citenamefont {Dane}, \citenamefont {Ivry},\ and\
  \citenamefont {Berggren}}]{zhao_superconducting-nanowire_2013}%
  \BibitemOpen
  \bibfield  {author} {\bibinfo {author} {\bibfnamefont {Q.}~\bibnamefont
  {Zhao}}, \bibinfo {author} {\bibfnamefont {A.}~\bibnamefont {McCaughan}},
  \bibinfo {author} {\bibfnamefont {F.}~\bibnamefont {Bellei}}, \bibinfo
  {author} {\bibfnamefont {F.}~\bibnamefont {Najafi}}, \bibinfo {author}
  {\bibfnamefont {D.}~\bibnamefont {De~Fazio}}, \bibinfo {author}
  {\bibfnamefont {A.}~\bibnamefont {Dane}}, \bibinfo {author} {\bibfnamefont
  {Y.}~\bibnamefont {Ivry}},\ and\ \bibinfo {author} {\bibfnamefont {K.~K.}\
  \bibnamefont {Berggren}},\ }\bibfield  {title} {\bibinfo {title}
  {Superconducting-nanowire single-photon-detector linear array},\ }\href
  {https://doi.org/10.1063/1.4823542} {\bibfield  {journal} {\bibinfo
  {journal} {Applied Physics Letters}\ }\textbf {\bibinfo {volume} {103}},\
  \bibinfo {pages} {142602} (\bibinfo {year} {2013})}\BibitemShut {NoStop}%
\bibitem [{\citenamefont {Rosenberg}\ \emph {et~al.}(2013)\citenamefont
  {Rosenberg}, \citenamefont {Kerman}, \citenamefont {Molnar},\ and\
  \citenamefont {Dauler}}]{rosenberg_high-speed_2013}%
  \BibitemOpen
  \bibfield  {author} {\bibinfo {author} {\bibfnamefont {D.}~\bibnamefont
  {Rosenberg}}, \bibinfo {author} {\bibfnamefont {A.~J.}\ \bibnamefont
  {Kerman}}, \bibinfo {author} {\bibfnamefont {R.~J.}\ \bibnamefont {Molnar}},\
  and\ \bibinfo {author} {\bibfnamefont {E.~A.}\ \bibnamefont {Dauler}},\
  }\bibfield  {title} {{\selectlanguage {english}\bibinfo {title} {High-speed
  and high-efficiency superconducting nanowire single photon detector array}},\
  }\href {https://doi.org/10.1364/OE.21.001440} {\bibfield  {journal} {\bibinfo
   {journal} {Optics Express}\ }\textbf {\bibinfo {volume} {21}},\ \bibinfo
  {pages} {1440} (\bibinfo {year} {2013})}\BibitemShut {NoStop}%
\bibitem [{\citenamefont {Verma}\ \emph {et~al.}(2014)\citenamefont {Verma},
  \citenamefont {Horansky}, \citenamefont {Marsili}, \citenamefont {Stern},
  \citenamefont {Shaw}, \citenamefont {Lita}, \citenamefont {Mirin},\ and\
  \citenamefont {Nam}}]{verma_four-pixel_2014}%
  \BibitemOpen
  \bibfield  {author} {\bibinfo {author} {\bibfnamefont {V.~B.}\ \bibnamefont
  {Verma}}, \bibinfo {author} {\bibfnamefont {R.}~\bibnamefont {Horansky}},
  \bibinfo {author} {\bibfnamefont {F.}~\bibnamefont {Marsili}}, \bibinfo
  {author} {\bibfnamefont {J.~A.}\ \bibnamefont {Stern}}, \bibinfo {author}
  {\bibfnamefont {M.~D.}\ \bibnamefont {Shaw}}, \bibinfo {author}
  {\bibfnamefont {A.~E.}\ \bibnamefont {Lita}}, \bibinfo {author}
  {\bibfnamefont {R.~P.}\ \bibnamefont {Mirin}},\ and\ \bibinfo {author}
  {\bibfnamefont {S.~W.}\ \bibnamefont {Nam}},\ }\bibfield  {title} {\bibinfo
  {title} {A four-pixel single-photon pulse-position array fabricated from
  {WSi} superconducting nanowire single-photon detectors},\ }\href
  {https://doi.org/10.1063/1.4864075} {\bibfield  {journal} {\bibinfo
  {journal} {Applied Physics Letters}\ }\textbf {\bibinfo {volume} {104}},\
  \bibinfo {pages} {051115} (\bibinfo {year} {2014})}\BibitemShut {NoStop}%
\bibitem [{\citenamefont {Allman}\ \emph {et~al.}(2015)\citenamefont {Allman},
  \citenamefont {Verma}, \citenamefont {Stevens}, \citenamefont {Gerrits},
  \citenamefont {Horansky}, \citenamefont {Lita}, \citenamefont {Marsili},
  \citenamefont {Beyer}, \citenamefont {Shaw}, \citenamefont {Kumor},
  \citenamefont {Mirin},\ and\ \citenamefont
  {Nam}}]{allman_near-infrared_2015}%
  \BibitemOpen
  \bibfield  {author} {\bibinfo {author} {\bibfnamefont {M.~S.}\ \bibnamefont
  {Allman}}, \bibinfo {author} {\bibfnamefont {V.~B.}\ \bibnamefont {Verma}},
  \bibinfo {author} {\bibfnamefont {M.}~\bibnamefont {Stevens}}, \bibinfo
  {author} {\bibfnamefont {T.}~\bibnamefont {Gerrits}}, \bibinfo {author}
  {\bibfnamefont {R.~D.}\ \bibnamefont {Horansky}}, \bibinfo {author}
  {\bibfnamefont {A.~E.}\ \bibnamefont {Lita}}, \bibinfo {author}
  {\bibfnamefont {F.}~\bibnamefont {Marsili}}, \bibinfo {author} {\bibfnamefont
  {A.}~\bibnamefont {Beyer}}, \bibinfo {author} {\bibfnamefont {M.~D.}\
  \bibnamefont {Shaw}}, \bibinfo {author} {\bibfnamefont {D.}~\bibnamefont
  {Kumor}}, \bibinfo {author} {\bibfnamefont {R.}~\bibnamefont {Mirin}},\ and\
  \bibinfo {author} {\bibfnamefont {S.~W.}\ \bibnamefont {Nam}},\ }\bibfield
  {title} {\bibinfo {title} {A near-infrared 64-pixel superconducting nanowire
  single photon detector array with integrated multiplexed readout},\ }\href
  {https://doi.org/10.1063/1.4921318} {\bibfield  {journal} {\bibinfo
  {journal} {Applied Physics Letters}\ }\textbf {\bibinfo {volume} {106}},\
  \bibinfo {pages} {192601} (\bibinfo {year} {2015})}\BibitemShut {NoStop}%
\bibitem [{\citenamefont {Najafi}\ \emph {et~al.}(2015)\citenamefont {Najafi},
  \citenamefont {Mower}, \citenamefont {Harris}, \citenamefont {Bellei},
  \citenamefont {Dane}, \citenamefont {Lee}, \citenamefont {Hu}, \citenamefont
  {Kharel}, \citenamefont {Marsili}, \citenamefont {Assefa}, \citenamefont
  {Berggren},\ and\ \citenamefont {Englund}}]{najafi_-chip_2015}%
  \BibitemOpen
  \bibfield  {author} {\bibinfo {author} {\bibfnamefont {F.}~\bibnamefont
  {Najafi}}, \bibinfo {author} {\bibfnamefont {J.}~\bibnamefont {Mower}},
  \bibinfo {author} {\bibfnamefont {N.~C.}\ \bibnamefont {Harris}}, \bibinfo
  {author} {\bibfnamefont {F.}~\bibnamefont {Bellei}}, \bibinfo {author}
  {\bibfnamefont {A.}~\bibnamefont {Dane}}, \bibinfo {author} {\bibfnamefont
  {C.}~\bibnamefont {Lee}}, \bibinfo {author} {\bibfnamefont {X.}~\bibnamefont
  {Hu}}, \bibinfo {author} {\bibfnamefont {P.}~\bibnamefont {Kharel}}, \bibinfo
  {author} {\bibfnamefont {F.}~\bibnamefont {Marsili}}, \bibinfo {author}
  {\bibfnamefont {S.}~\bibnamefont {Assefa}}, \bibinfo {author} {\bibfnamefont
  {K.~K.}\ \bibnamefont {Berggren}},\ and\ \bibinfo {author} {\bibfnamefont
  {D.}~\bibnamefont {Englund}},\ }\bibfield  {title} {{\selectlanguage
  {english}\bibinfo {title} {On-chip detection of non-classical light by
  scalable integration of single-photon detectors}},\ }\href
  {https://doi.org/10.1038/ncomms6873} {\bibfield  {journal} {\bibinfo
  {journal} {Nature Communications}\ }\textbf {\bibinfo {volume} {6}},\
  \bibinfo {pages} {5873} (\bibinfo {year} {2015})}\BibitemShut {NoStop}%
\bibitem [{\citenamefont {Chen}\ \emph {et~al.}(2018)\citenamefont {Chen},
  \citenamefont {Zhang}, \citenamefont {Zhang}, \citenamefont {Ge},
  \citenamefont {Xu}, \citenamefont {Wu}, \citenamefont {Tu}, \citenamefont
  {Jia}, \citenamefont {Kang}, \citenamefont {Chen},\ and\ \citenamefont
  {Wu}}]{chen_16-pixel_2018}%
  \BibitemOpen
  \bibfield  {author} {\bibinfo {author} {\bibfnamefont {Q.}~\bibnamefont
  {Chen}}, \bibinfo {author} {\bibfnamefont {B.}~\bibnamefont {Zhang}},
  \bibinfo {author} {\bibfnamefont {L.}~\bibnamefont {Zhang}}, \bibinfo
  {author} {\bibfnamefont {R.}~\bibnamefont {Ge}}, \bibinfo {author}
  {\bibfnamefont {R.}~\bibnamefont {Xu}}, \bibinfo {author} {\bibfnamefont
  {Y.}~\bibnamefont {Wu}}, \bibinfo {author} {\bibfnamefont {X.}~\bibnamefont
  {Tu}}, \bibinfo {author} {\bibfnamefont {X.}~\bibnamefont {Jia}}, \bibinfo
  {author} {\bibfnamefont {L.}~\bibnamefont {Kang}}, \bibinfo {author}
  {\bibfnamefont {J.}~\bibnamefont {Chen}},\ and\ \bibinfo {author}
  {\bibfnamefont {P.}~\bibnamefont {Wu}},\ }\bibfield  {title}
  {{\selectlanguage {english}\bibinfo {title} {A 16-pixel {NbN} nanowire single
  photon detector coupled with 300 micrometer fiber}},\ }\href
  {https://arxiv.org/abs/1811.09779v1} {\bibfield  {journal} {\bibinfo
  {journal} {arXiv preprint arXiv:1811.09779}\ } (\bibinfo {year}
  {2018})}\BibitemShut {NoStop}%
\bibitem [{\citenamefont {Tao}\ \emph {et~al.}(2019)\citenamefont {Tao},
  \citenamefont {Chen}, \citenamefont {Chen}, \citenamefont {Wang},
  \citenamefont {Li}, \citenamefont {Tu}, \citenamefont {Jia}, \citenamefont
  {Zhao}, \citenamefont {Zhang}, \citenamefont {Kang} \emph
  {et~al.}}]{tao2019high}%
  \BibitemOpen
  \bibfield  {author} {\bibinfo {author} {\bibfnamefont {X.}~\bibnamefont
  {Tao}}, \bibinfo {author} {\bibfnamefont {S.}~\bibnamefont {Chen}}, \bibinfo
  {author} {\bibfnamefont {Y.}~\bibnamefont {Chen}}, \bibinfo {author}
  {\bibfnamefont {L.}~\bibnamefont {Wang}}, \bibinfo {author} {\bibfnamefont
  {X.}~\bibnamefont {Li}}, \bibinfo {author} {\bibfnamefont {X.}~\bibnamefont
  {Tu}}, \bibinfo {author} {\bibfnamefont {X.}~\bibnamefont {Jia}}, \bibinfo
  {author} {\bibfnamefont {Q.}~\bibnamefont {Zhao}}, \bibinfo {author}
  {\bibfnamefont {L.}~\bibnamefont {Zhang}}, \bibinfo {author} {\bibfnamefont
  {L.}~\bibnamefont {Kang}}, \emph {et~al.},\ }\bibfield  {title} {\bibinfo
  {title} {A high speed and high efficiency superconducting photon number
  resolving detector},\ }\href@noop {} {\bibfield  {journal} {\bibinfo
  {journal} {Superconductor Science and Technology}\ }\textbf {\bibinfo
  {volume} {32}},\ \bibinfo {pages} {064002} (\bibinfo {year}
  {2019})}\BibitemShut {NoStop}%
\bibitem [{\citenamefont {Wollman}\ \emph {et~al.}(2019)\citenamefont
  {Wollman}, \citenamefont {Verma}, \citenamefont {Lita}, \citenamefont {Farr},
  \citenamefont {Shaw}, \citenamefont {Mirin},\ and\ \citenamefont
  {Nam}}]{wollman2019kilopixel}%
  \BibitemOpen
  \bibfield  {author} {\bibinfo {author} {\bibfnamefont {E.~E.}\ \bibnamefont
  {Wollman}}, \bibinfo {author} {\bibfnamefont {V.~B.}\ \bibnamefont {Verma}},
  \bibinfo {author} {\bibfnamefont {A.~E.}\ \bibnamefont {Lita}}, \bibinfo
  {author} {\bibfnamefont {W.~H.}\ \bibnamefont {Farr}}, \bibinfo {author}
  {\bibfnamefont {M.~D.}\ \bibnamefont {Shaw}}, \bibinfo {author}
  {\bibfnamefont {R.~P.}\ \bibnamefont {Mirin}},\ and\ \bibinfo {author}
  {\bibfnamefont {S.~W.}\ \bibnamefont {Nam}},\ }\bibfield  {title} {\bibinfo
  {title} {Kilopixel array of superconducting nanowire single-photon
  detectors},\ }\href@noop {} {\bibfield  {journal} {\bibinfo  {journal}
  {Optics Express}\ }\textbf {\bibinfo {volume} {27}},\ \bibinfo {pages}
  {35279} (\bibinfo {year} {2019})}\BibitemShut {NoStop}%
\bibitem [{\citenamefont {Miki}\ \emph {et~al.}(2014)\citenamefont {Miki},
  \citenamefont {Yamashita}, \citenamefont {Wang},\ and\ \citenamefont
  {Terai}}]{miki_64-pixel_2014}%
  \BibitemOpen
  \bibfield  {author} {\bibinfo {author} {\bibfnamefont {S.}~\bibnamefont
  {Miki}}, \bibinfo {author} {\bibfnamefont {T.}~\bibnamefont {Yamashita}},
  \bibinfo {author} {\bibfnamefont {Z.}~\bibnamefont {Wang}},\ and\ \bibinfo
  {author} {\bibfnamefont {H.}~\bibnamefont {Terai}},\ }\bibfield  {title}
  {{\selectlanguage {english}\bibinfo {title} {A 64-pixel {NbTiN}
  superconducting nanowire single-photon detector array for spatially resolved
  photon detection}},\ }\href {https://doi.org/10.1364/OE.22.007811} {\bibfield
   {journal} {\bibinfo  {journal} {Optics Express}\ }\textbf {\bibinfo {volume}
  {22}},\ \bibinfo {pages} {7811} (\bibinfo {year} {2014})}\BibitemShut
  {NoStop}%
\bibitem [{\citenamefont {Moreau}\ \emph {et~al.}(2019)\citenamefont {Moreau},
  \citenamefont {Toninelli}, \citenamefont {Gregory},\ and\ \citenamefont
  {Padgett}}]{moreau2019imaging}%
  \BibitemOpen
  \bibfield  {author} {\bibinfo {author} {\bibfnamefont {P.-A.}\ \bibnamefont
  {Moreau}}, \bibinfo {author} {\bibfnamefont {E.}~\bibnamefont {Toninelli}},
  \bibinfo {author} {\bibfnamefont {T.}~\bibnamefont {Gregory}},\ and\ \bibinfo
  {author} {\bibfnamefont {M.~J.}\ \bibnamefont {Padgett}},\ }\bibfield
  {title} {\bibinfo {title} {Imaging with quantum states of light},\
  }\href@noop {} {\bibfield  {journal} {\bibinfo  {journal} {Nature Reviews
  Physics}\ }\textbf {\bibinfo {volume} {1}},\ \bibinfo {pages} {367} (\bibinfo
  {year} {2019})}\BibitemShut {NoStop}%
\bibitem [{\citenamefont {Shaw}\ \emph {et~al.}(2015)\citenamefont {Shaw},
  \citenamefont {Marsili}, \citenamefont {Beyer}, \citenamefont {Stern},
  \citenamefont {Resta}, \citenamefont {Ravindran}, \citenamefont {Chang},
  \citenamefont {Bardin}, \citenamefont {Russell}, \citenamefont {Gin},
  \citenamefont {Patawaran}, \citenamefont {Verma}, \citenamefont {Mirin},
  \citenamefont {Nam},\ and\ \citenamefont {Farr}}]{shaw_arrays_2015}%
  \BibitemOpen
  \bibfield  {author} {\bibinfo {author} {\bibfnamefont {M.~D.}\ \bibnamefont
  {Shaw}}, \bibinfo {author} {\bibfnamefont {F.}~\bibnamefont {Marsili}},
  \bibinfo {author} {\bibfnamefont {A.~D.}\ \bibnamefont {Beyer}}, \bibinfo
  {author} {\bibfnamefont {J.~A.}\ \bibnamefont {Stern}}, \bibinfo {author}
  {\bibfnamefont {G.~V.}\ \bibnamefont {Resta}}, \bibinfo {author}
  {\bibfnamefont {P.}~\bibnamefont {Ravindran}}, \bibinfo {author}
  {\bibfnamefont {S.}~\bibnamefont {Chang}}, \bibinfo {author} {\bibfnamefont
  {J.}~\bibnamefont {Bardin}}, \bibinfo {author} {\bibfnamefont {D.~S.}\
  \bibnamefont {Russell}}, \bibinfo {author} {\bibfnamefont {J.~W.}\
  \bibnamefont {Gin}}, \bibinfo {author} {\bibfnamefont {F.~D.}\ \bibnamefont
  {Patawaran}}, \bibinfo {author} {\bibfnamefont {V.~B.}\ \bibnamefont
  {Verma}}, \bibinfo {author} {\bibfnamefont {R.~P.}\ \bibnamefont {Mirin}},
  \bibinfo {author} {\bibfnamefont {S.~W.}\ \bibnamefont {Nam}},\ and\ \bibinfo
  {author} {\bibfnamefont {W.~H.}\ \bibnamefont {Farr}},\ }\bibfield  {title}
  {{\selectlanguage {english}\bibinfo {title} {Arrays of {WSi}
  {Superconducting} {Nanowire} {Single} {Photon} {Detectors} for {Deep}-{Space}
  {Optical} {Communications}}},\ }in\ \href
  {https://doi.org/10.1364/CLEO_AT.2015.JTh2A.68} {{\selectlanguage
  {english}\emph {\bibinfo {booktitle} {2015 Conference on Lasers and
  Electro-Optics (CLEO). IEEE}}}}\ (\bibinfo {year} {2015})\BibitemShut
  {NoStop}%
\bibitem [{\citenamefont {Cahall}\ \emph {et~al.}(2017)\citenamefont {Cahall},
  \citenamefont {Nicolich}, \citenamefont {Islam}, \citenamefont {Lafyatis},
  \citenamefont {Miller}, \citenamefont {Gauthier},\ and\ \citenamefont
  {Kim}}]{cahall2017multi}%
  \BibitemOpen
  \bibfield  {author} {\bibinfo {author} {\bibfnamefont {C.}~\bibnamefont
  {Cahall}}, \bibinfo {author} {\bibfnamefont {K.~L.}\ \bibnamefont
  {Nicolich}}, \bibinfo {author} {\bibfnamefont {N.~T.}\ \bibnamefont {Islam}},
  \bibinfo {author} {\bibfnamefont {G.~P.}\ \bibnamefont {Lafyatis}}, \bibinfo
  {author} {\bibfnamefont {A.~J.}\ \bibnamefont {Miller}}, \bibinfo {author}
  {\bibfnamefont {D.~J.}\ \bibnamefont {Gauthier}},\ and\ \bibinfo {author}
  {\bibfnamefont {J.}~\bibnamefont {Kim}},\ }\bibfield  {title} {\bibinfo
  {title} {Multi-photon detection using a conventional superconducting nanowire
  single-photon detector},\ }\href@noop {} {\bibfield  {journal} {\bibinfo
  {journal} {Optica}\ }\textbf {\bibinfo {volume} {4}},\ \bibinfo {pages}
  {1534} (\bibinfo {year} {2017})}\BibitemShut {NoStop}%
\bibitem [{\citenamefont {Zhu}\ \emph {et~al.}(2018)\citenamefont {Zhu},
  \citenamefont {Zhao}, \citenamefont {Choi}, \citenamefont {Lu}, \citenamefont
  {Dane}, \citenamefont {Englund},\ and\ \citenamefont
  {Berggren}}]{zhu2018scalable}%
  \BibitemOpen
  \bibfield  {author} {\bibinfo {author} {\bibfnamefont {D.}~\bibnamefont
  {Zhu}}, \bibinfo {author} {\bibfnamefont {Q.-Y.}\ \bibnamefont {Zhao}},
  \bibinfo {author} {\bibfnamefont {H.}~\bibnamefont {Choi}}, \bibinfo {author}
  {\bibfnamefont {T.-J.}\ \bibnamefont {Lu}}, \bibinfo {author} {\bibfnamefont
  {A.~E.}\ \bibnamefont {Dane}}, \bibinfo {author} {\bibfnamefont
  {D.}~\bibnamefont {Englund}},\ and\ \bibinfo {author} {\bibfnamefont {K.~K.}\
  \bibnamefont {Berggren}},\ }\bibfield  {title} {\bibinfo {title} {A scalable
  multi-photon coincidence detector based on superconducting nanowires},\
  }\href@noop {} {\bibfield  {journal} {\bibinfo  {journal} {Nature
  Nanotechnol.}\ }\textbf {\bibinfo {volume} {13}},\ \bibinfo {pages} {596}
  (\bibinfo {year} {2018})}\BibitemShut {NoStop}%
\bibitem [{\citenamefont {Zhu}\ \emph {et~al.}(2020)\citenamefont {Zhu},
  \citenamefont {Colangelo}, \citenamefont {Chen}, \citenamefont {Korzh},
  \citenamefont {Wong}, \citenamefont {Shaw},\ and\ \citenamefont
  {Berggren}}]{zhu2020resolving}%
  \BibitemOpen
  \bibfield  {author} {\bibinfo {author} {\bibfnamefont {D.}~\bibnamefont
  {Zhu}}, \bibinfo {author} {\bibfnamefont {M.}~\bibnamefont {Colangelo}},
  \bibinfo {author} {\bibfnamefont {C.}~\bibnamefont {Chen}}, \bibinfo {author}
  {\bibfnamefont {B.~A.}\ \bibnamefont {Korzh}}, \bibinfo {author}
  {\bibfnamefont {F.~N.}\ \bibnamefont {Wong}}, \bibinfo {author}
  {\bibfnamefont {M.~D.}\ \bibnamefont {Shaw}},\ and\ \bibinfo {author}
  {\bibfnamefont {K.~K.}\ \bibnamefont {Berggren}},\ }\bibfield  {title}
  {\bibinfo {title} {Resolving photon numbers using a superconducting nanowire
  with impedance-matching taper},\ }\href@noop {} {\bibfield  {journal}
  {\bibinfo  {journal} {Nano Letters}\ }\textbf {\bibinfo {volume} {20}},\
  \bibinfo {pages} {3858} (\bibinfo {year} {2020})}\BibitemShut {NoStop}%
\bibitem [{\citenamefont {Zou}\ \emph {et~al.}(2020)\citenamefont {Zou},
  \citenamefont {Meng}, \citenamefont {Wang},\ and\ \citenamefont
  {Hu}}]{zou2020superconducting}%
  \BibitemOpen
  \bibfield  {author} {\bibinfo {author} {\bibfnamefont {K.}~\bibnamefont
  {Zou}}, \bibinfo {author} {\bibfnamefont {Y.}~\bibnamefont {Meng}}, \bibinfo
  {author} {\bibfnamefont {Z.}~\bibnamefont {Wang}},\ and\ \bibinfo {author}
  {\bibfnamefont {X.}~\bibnamefont {Hu}},\ }\bibfield  {title} {\bibinfo
  {title} {Superconducting nanowire multi-photon detectors enabled by current
  reservoirs},\ }\href@noop {} {\bibfield  {journal} {\bibinfo  {journal}
  {Photonics Research}\ }\textbf {\bibinfo {volume} {8}},\ \bibinfo {pages}
  {601} (\bibinfo {year} {2020})}\BibitemShut {NoStop}%
\bibitem [{\citenamefont {Natarajan}\ \emph {et~al.}(2013)\citenamefont
  {Natarajan}, \citenamefont {Zhang}, \citenamefont {Coldenstrodt-Ronge},
  \citenamefont {Donati}, \citenamefont {Dorenbos}, \citenamefont {Zwiller},
  \citenamefont {Walmsley},\ and\ \citenamefont
  {Hadfield}}]{natarajan2013quantum}%
  \BibitemOpen
  \bibfield  {author} {\bibinfo {author} {\bibfnamefont {C.~M.}\ \bibnamefont
  {Natarajan}}, \bibinfo {author} {\bibfnamefont {L.}~\bibnamefont {Zhang}},
  \bibinfo {author} {\bibfnamefont {H.}~\bibnamefont {Coldenstrodt-Ronge}},
  \bibinfo {author} {\bibfnamefont {G.}~\bibnamefont {Donati}}, \bibinfo
  {author} {\bibfnamefont {S.~N.}\ \bibnamefont {Dorenbos}}, \bibinfo {author}
  {\bibfnamefont {V.}~\bibnamefont {Zwiller}}, \bibinfo {author} {\bibfnamefont
  {I.~A.}\ \bibnamefont {Walmsley}},\ and\ \bibinfo {author} {\bibfnamefont
  {R.~H.}\ \bibnamefont {Hadfield}},\ }\bibfield  {title} {\bibinfo {title}
  {Quantum detector tomography of a time-multiplexed superconducting nanowire
  single-photon detector at telecom wavelengths},\ }\href@noop {} {\bibfield
  {journal} {\bibinfo  {journal} {Opt. Express}\ }\textbf {\bibinfo {volume}
  {21}},\ \bibinfo {pages} {893} (\bibinfo {year} {2013})}\BibitemShut
  {NoStop}%
\bibitem [{\citenamefont {Tiedau}\ \emph {et~al.}(2019)\citenamefont {Tiedau},
  \citenamefont {Meyer-Scott}, \citenamefont {Nitsche}, \citenamefont
  {Barkhofen}, \citenamefont {Bartley},\ and\ \citenamefont
  {Silberhorn}}]{tiedau2019high}%
  \BibitemOpen
  \bibfield  {author} {\bibinfo {author} {\bibfnamefont {J.}~\bibnamefont
  {Tiedau}}, \bibinfo {author} {\bibfnamefont {E.}~\bibnamefont {Meyer-Scott}},
  \bibinfo {author} {\bibfnamefont {T.}~\bibnamefont {Nitsche}}, \bibinfo
  {author} {\bibfnamefont {S.}~\bibnamefont {Barkhofen}}, \bibinfo {author}
  {\bibfnamefont {T.~J.}\ \bibnamefont {Bartley}},\ and\ \bibinfo {author}
  {\bibfnamefont {C.}~\bibnamefont {Silberhorn}},\ }\bibfield  {title}
  {\bibinfo {title} {A high dynamic range optical detector for measuring single
  photons and bright light},\ }\href@noop {} {\bibfield  {journal} {\bibinfo
  {journal} {Opt Express}\ }\textbf {\bibinfo {volume} {27}},\ \bibinfo {pages}
  {1} (\bibinfo {year} {2019})}\BibitemShut {NoStop}%
\bibitem [{\citenamefont {Lundeen}\ \emph {et~al.}(2009)\citenamefont
  {Lundeen}, \citenamefont {Feito}, \citenamefont {Coldenstrodt-Ronge},
  \citenamefont {Pregnell}, \citenamefont {Silberhorn}, \citenamefont {Ralph},
  \citenamefont {Eisert}, \citenamefont {Plenio},\ and\ \citenamefont
  {Walmsley}}]{lundeen2009tomography}%
  \BibitemOpen
  \bibfield  {author} {\bibinfo {author} {\bibfnamefont {J.}~\bibnamefont
  {Lundeen}}, \bibinfo {author} {\bibfnamefont {A.}~\bibnamefont {Feito}},
  \bibinfo {author} {\bibfnamefont {H.}~\bibnamefont {Coldenstrodt-Ronge}},
  \bibinfo {author} {\bibfnamefont {K.}~\bibnamefont {Pregnell}}, \bibinfo
  {author} {\bibfnamefont {C.}~\bibnamefont {Silberhorn}}, \bibinfo {author}
  {\bibfnamefont {T.}~\bibnamefont {Ralph}}, \bibinfo {author} {\bibfnamefont
  {J.}~\bibnamefont {Eisert}}, \bibinfo {author} {\bibfnamefont
  {M.}~\bibnamefont {Plenio}},\ and\ \bibinfo {author} {\bibfnamefont
  {I.}~\bibnamefont {Walmsley}},\ }\bibfield  {title} {\bibinfo {title}
  {Tomography of quantum detectors},\ }\href@noop {} {\bibfield  {journal}
  {\bibinfo  {journal} {Nature Physics}\ }\textbf {\bibinfo {volume} {5}},\
  \bibinfo {pages} {27} (\bibinfo {year} {2009})}\BibitemShut {NoStop}%
\bibitem [{\citenamefont {Feito}\ \emph {et~al.}(2009)\citenamefont {Feito},
  \citenamefont {Lundeen}, \citenamefont {Coldenstrodt-Ronge}, \citenamefont
  {Eisert}, \citenamefont {Plenio},\ and\ \citenamefont
  {Walmsley}}]{feito2009measuring}%
  \BibitemOpen
  \bibfield  {author} {\bibinfo {author} {\bibfnamefont {A.}~\bibnamefont
  {Feito}}, \bibinfo {author} {\bibfnamefont {J.}~\bibnamefont {Lundeen}},
  \bibinfo {author} {\bibfnamefont {H.}~\bibnamefont {Coldenstrodt-Ronge}},
  \bibinfo {author} {\bibfnamefont {J.}~\bibnamefont {Eisert}}, \bibinfo
  {author} {\bibfnamefont {M.~B.}\ \bibnamefont {Plenio}},\ and\ \bibinfo
  {author} {\bibfnamefont {I.~A.}\ \bibnamefont {Walmsley}},\ }\bibfield
  {title} {\bibinfo {title} {Measuring measurement: theory and practice},\
  }\href@noop {} {\bibfield  {journal} {\bibinfo  {journal} {New Journal of
  Physics}\ }\textbf {\bibinfo {volume} {11}},\ \bibinfo {pages} {093038}
  (\bibinfo {year} {2009})}\BibitemShut {NoStop}%
\bibitem [{\citenamefont {Piacentini}\ \emph {et~al.}(2015)\citenamefont
  {Piacentini}, \citenamefont {Levi}, \citenamefont {Avella}, \citenamefont
  {L{\'o}pez}, \citenamefont {K{\"u}ck}, \citenamefont {Polyakov},
  \citenamefont {Degiovanni}, \citenamefont {Brida},\ and\ \citenamefont
  {Genovese}}]{piacentini2015positive}%
  \BibitemOpen
  \bibfield  {author} {\bibinfo {author} {\bibfnamefont {F.}~\bibnamefont
  {Piacentini}}, \bibinfo {author} {\bibfnamefont {M.}~\bibnamefont {Levi}},
  \bibinfo {author} {\bibfnamefont {A.}~\bibnamefont {Avella}}, \bibinfo
  {author} {\bibfnamefont {M.}~\bibnamefont {L{\'o}pez}}, \bibinfo {author}
  {\bibfnamefont {S.}~\bibnamefont {K{\"u}ck}}, \bibinfo {author}
  {\bibfnamefont {S.}~\bibnamefont {Polyakov}}, \bibinfo {author}
  {\bibfnamefont {I.~P.}\ \bibnamefont {Degiovanni}}, \bibinfo {author}
  {\bibfnamefont {G.}~\bibnamefont {Brida}},\ and\ \bibinfo {author}
  {\bibfnamefont {M.}~\bibnamefont {Genovese}},\ }\bibfield  {title} {\bibinfo
  {title} {Positive operator-valued measure reconstruction of a beam-splitter
  tree-based photon-number-resolving detector},\ }\href@noop {} {\bibfield
  {journal} {\bibinfo  {journal} {Opt. Lett.}\ }\textbf {\bibinfo {volume}
  {40}},\ \bibinfo {pages} {1548} (\bibinfo {year} {2015})}\BibitemShut
  {NoStop}%
\bibitem [{\citenamefont {Humphreys}\ \emph {et~al.}(2015)\citenamefont
  {Humphreys}, \citenamefont {Metcalf}, \citenamefont {Gerrits}, \citenamefont
  {Hiemstra}, \citenamefont {Lita}, \citenamefont {Nunn}, \citenamefont {Nam},
  \citenamefont {Datta}, \citenamefont {Kolthammer},\ and\ \citenamefont
  {Walmsley}}]{humphreys2015tomography}%
  \BibitemOpen
  \bibfield  {author} {\bibinfo {author} {\bibfnamefont {P.~C.}\ \bibnamefont
  {Humphreys}}, \bibinfo {author} {\bibfnamefont {B.~J.}\ \bibnamefont
  {Metcalf}}, \bibinfo {author} {\bibfnamefont {T.}~\bibnamefont {Gerrits}},
  \bibinfo {author} {\bibfnamefont {T.}~\bibnamefont {Hiemstra}}, \bibinfo
  {author} {\bibfnamefont {A.~E.}\ \bibnamefont {Lita}}, \bibinfo {author}
  {\bibfnamefont {J.}~\bibnamefont {Nunn}}, \bibinfo {author} {\bibfnamefont
  {S.~W.}\ \bibnamefont {Nam}}, \bibinfo {author} {\bibfnamefont
  {A.}~\bibnamefont {Datta}}, \bibinfo {author} {\bibfnamefont {W.~S.}\
  \bibnamefont {Kolthammer}},\ and\ \bibinfo {author} {\bibfnamefont {I.~A.}\
  \bibnamefont {Walmsley}},\ }\bibfield  {title} {\bibinfo {title} {Tomography
  of photon-number resolving continuous-output detectors},\ }\href@noop {}
  {\bibfield  {journal} {\bibinfo  {journal} {New Journal of Physics}\ }\textbf
  {\bibinfo {volume} {17}},\ \bibinfo {pages} {103044} (\bibinfo {year}
  {2015})}\BibitemShut {NoStop}%
\bibitem [{\citenamefont {Zhang}\ \emph {et~al.}(2012)\citenamefont {Zhang},
  \citenamefont {Coldenstrodt-Ronge}, \citenamefont {Datta}, \citenamefont
  {Puentes}, \citenamefont {Lundeen}, \citenamefont {Jin}, \citenamefont
  {Smith}, \citenamefont {Plenio},\ and\ \citenamefont
  {Walmsley}}]{zhang2012mapping}%
  \BibitemOpen
  \bibfield  {author} {\bibinfo {author} {\bibfnamefont {L.}~\bibnamefont
  {Zhang}}, \bibinfo {author} {\bibfnamefont {H.~B.}\ \bibnamefont
  {Coldenstrodt-Ronge}}, \bibinfo {author} {\bibfnamefont {A.}~\bibnamefont
  {Datta}}, \bibinfo {author} {\bibfnamefont {G.}~\bibnamefont {Puentes}},
  \bibinfo {author} {\bibfnamefont {J.~S.}\ \bibnamefont {Lundeen}}, \bibinfo
  {author} {\bibfnamefont {X.-M.}\ \bibnamefont {Jin}}, \bibinfo {author}
  {\bibfnamefont {B.~J.}\ \bibnamefont {Smith}}, \bibinfo {author}
  {\bibfnamefont {M.~B.}\ \bibnamefont {Plenio}},\ and\ \bibinfo {author}
  {\bibfnamefont {I.~A.}\ \bibnamefont {Walmsley}},\ }\bibfield  {title}
  {\bibinfo {title} {Mapping coherence in measurement via full quantum
  tomography of a hybrid optical detector},\ }\href@noop {} {\bibfield
  {journal} {\bibinfo  {journal} {Nature Photonics}\ }\textbf {\bibinfo
  {volume} {6}},\ \bibinfo {pages} {364} (\bibinfo {year} {2012})}\BibitemShut
  {NoStop}%
\bibitem [{\citenamefont {Cooper}\ \emph {et~al.}(2014)\citenamefont {Cooper},
  \citenamefont {Karpi{\'n}ski},\ and\ \citenamefont
  {Smith}}]{cooper2014local}%
  \BibitemOpen
  \bibfield  {author} {\bibinfo {author} {\bibfnamefont {M.}~\bibnamefont
  {Cooper}}, \bibinfo {author} {\bibfnamefont {M.}~\bibnamefont
  {Karpi{\'n}ski}},\ and\ \bibinfo {author} {\bibfnamefont {B.~J.}\
  \bibnamefont {Smith}},\ }\bibfield  {title} {\bibinfo {title} {Local mapping
  of detector response for reliable quantum state estimation},\ }\href@noop {}
  {\bibfield  {journal} {\bibinfo  {journal} {Nat. Commun.}\ }\textbf {\bibinfo
  {volume} {5}},\ \bibinfo {pages} {4332} (\bibinfo {year} {2014})}\BibitemShut
  {NoStop}%
\bibitem [{\citenamefont {Afek}\ \emph {et~al.}(2009)\citenamefont {Afek},
  \citenamefont {Natan}, \citenamefont {Ambar},\ and\ \citenamefont
  {Silberberg}}]{afek2009quantum}%
  \BibitemOpen
  \bibfield  {author} {\bibinfo {author} {\bibfnamefont {I.}~\bibnamefont
  {Afek}}, \bibinfo {author} {\bibfnamefont {A.}~\bibnamefont {Natan}},
  \bibinfo {author} {\bibfnamefont {O.}~\bibnamefont {Ambar}},\ and\ \bibinfo
  {author} {\bibfnamefont {Y.}~\bibnamefont {Silberberg}},\ }\bibfield  {title}
  {\bibinfo {title} {Quantum state measurements using multipixel photon
  detectors},\ }\href@noop {} {\bibfield  {journal} {\bibinfo  {journal}
  {Physical Review A}\ }\textbf {\bibinfo {volume} {79}},\ \bibinfo {pages}
  {043830} (\bibinfo {year} {2009})}\BibitemShut {NoStop}%
\bibitem [{\citenamefont {Nehra}\ \emph {et~al.}(2020)\citenamefont {Nehra},
  \citenamefont {Chang}, \citenamefont {Yu}, \citenamefont {Beling},\ and\
  \citenamefont {Pfister}}]{nehra2020photon}%
  \BibitemOpen
  \bibfield  {author} {\bibinfo {author} {\bibfnamefont {R.}~\bibnamefont
  {Nehra}}, \bibinfo {author} {\bibfnamefont {C.-H.}\ \bibnamefont {Chang}},
  \bibinfo {author} {\bibfnamefont {Q.}~\bibnamefont {Yu}}, \bibinfo {author}
  {\bibfnamefont {A.}~\bibnamefont {Beling}},\ and\ \bibinfo {author}
  {\bibfnamefont {O.}~\bibnamefont {Pfister}},\ }\bibfield  {title} {\bibinfo
  {title} {Photon-number-resolving segmented detectors based on single-photon
  avalanche-photodiodes},\ }\href@noop {} {\bibfield  {journal} {\bibinfo
  {journal} {Optics Express}\ }\textbf {\bibinfo {volume} {28}},\ \bibinfo
  {pages} {3660} (\bibinfo {year} {2020})}\BibitemShut {NoStop}%
\bibitem [{\citenamefont {Eraerds}\ \emph {et~al.}(2007)\citenamefont
  {Eraerds}, \citenamefont {Legr{\'e}}, \citenamefont {Rochas}, \citenamefont
  {Zbinden},\ and\ \citenamefont {Gisin}}]{eraerds2007sipm}%
  \BibitemOpen
  \bibfield  {author} {\bibinfo {author} {\bibfnamefont {P.}~\bibnamefont
  {Eraerds}}, \bibinfo {author} {\bibfnamefont {M.}~\bibnamefont {Legr{\'e}}},
  \bibinfo {author} {\bibfnamefont {A.}~\bibnamefont {Rochas}}, \bibinfo
  {author} {\bibfnamefont {H.}~\bibnamefont {Zbinden}},\ and\ \bibinfo {author}
  {\bibfnamefont {N.}~\bibnamefont {Gisin}},\ }\bibfield  {title} {\bibinfo
  {title} {Sipm for fast photon-counting and multiphoton detection},\
  }\href@noop {} {\bibfield  {journal} {\bibinfo  {journal} {Optics Express}\
  }\textbf {\bibinfo {volume} {15}},\ \bibinfo {pages} {14539} (\bibinfo {year}
  {2007})}\BibitemShut {NoStop}%
\bibitem [{\citenamefont {Akiba}\ \emph {et~al.}(2008)\citenamefont {Akiba},
  \citenamefont {Tsujino}, \citenamefont {Sato},\ and\ \citenamefont
  {Sasaki}}]{akiba2008multipixel}%
  \BibitemOpen
  \bibfield  {author} {\bibinfo {author} {\bibfnamefont {M.}~\bibnamefont
  {Akiba}}, \bibinfo {author} {\bibfnamefont {K.}~\bibnamefont {Tsujino}},
  \bibinfo {author} {\bibfnamefont {K.}~\bibnamefont {Sato}},\ and\ \bibinfo
  {author} {\bibfnamefont {M.}~\bibnamefont {Sasaki}},\ }\bibfield  {title}
  {\bibinfo {title} {A multipixel silicon apd with ultralow dark count rate at
  liquid nitrogen temperature},\ }\href@noop {} {\bibfield  {journal} {\bibinfo
   {journal} {arXiv preprint arXiv:0812.0634}\ } (\bibinfo {year}
  {2008})}\BibitemShut {NoStop}%
\bibitem [{\citenamefont {Kalashnikov}\ \emph {et~al.}(2012)\citenamefont
  {Kalashnikov}, \citenamefont {Tan},\ and\ \citenamefont
  {Krivitsky}}]{kalashnikov2012crosstalk}%
  \BibitemOpen
  \bibfield  {author} {\bibinfo {author} {\bibfnamefont {D.~A.}\ \bibnamefont
  {Kalashnikov}}, \bibinfo {author} {\bibfnamefont {S.-H.}\ \bibnamefont
  {Tan}},\ and\ \bibinfo {author} {\bibfnamefont {L.~A.}\ \bibnamefont
  {Krivitsky}},\ }\bibfield  {title} {\bibinfo {title} {Crosstalk calibration
  of multi-pixel photon counters using coherent states},\ }\href@noop {}
  {\bibfield  {journal} {\bibinfo  {journal} {Opt. Express}\ }\textbf {\bibinfo
  {volume} {20}},\ \bibinfo {pages} {5044} (\bibinfo {year}
  {2012})}\BibitemShut {NoStop}%
\bibitem [{\citenamefont {Golub}\ and\ \citenamefont
  {Van~Loan}(2013)}]{golub2013matrix}%
  \BibitemOpen
  \bibfield  {author} {\bibinfo {author} {\bibfnamefont {G.}~\bibnamefont
  {Golub}}\ and\ \bibinfo {author} {\bibfnamefont {C.}~\bibnamefont
  {Van~Loan}},\ }\href {https://books.google.de/books?id=X5YfsuCWpxMC} {\emph
  {\bibinfo {title} {Matrix Computations}}},\ Johns Hopkins Studies in the
  Mathematical Sciences\ (\bibinfo  {publisher} {Johns Hopkins University
  Press},\ \bibinfo {year} {2013})\BibitemShut {NoStop}%
\bibitem [{\citenamefont {Diamond}\ and\ \citenamefont {Boyd}(2016)}]{cvxpy}%
  \BibitemOpen
  \bibfield  {author} {\bibinfo {author} {\bibfnamefont {S.}~\bibnamefont
  {Diamond}}\ and\ \bibinfo {author} {\bibfnamefont {S.}~\bibnamefont {Boyd}},\
  }\bibfield  {title} {\bibinfo {title} {{CVXPY}: A {P}ython-embedded modeling
  language for convex optimization},\ }\href@noop {} {\bibfield  {journal}
  {\bibinfo  {journal} {Journal of Machine Learning Research}\ }\textbf
  {\bibinfo {volume} {17}},\ \bibinfo {pages} {1} (\bibinfo {year}
  {2016})}\BibitemShut {NoStop}%
\bibitem [{\citenamefont {Agrawal}\ \emph {et~al.}(2018)\citenamefont
  {Agrawal}, \citenamefont {Verschueren}, \citenamefont {Diamond},\ and\
  \citenamefont {Boyd}}]{cvxpy_rewriting}%
  \BibitemOpen
  \bibfield  {author} {\bibinfo {author} {\bibfnamefont {A.}~\bibnamefont
  {Agrawal}}, \bibinfo {author} {\bibfnamefont {R.}~\bibnamefont
  {Verschueren}}, \bibinfo {author} {\bibfnamefont {S.}~\bibnamefont
  {Diamond}},\ and\ \bibinfo {author} {\bibfnamefont {S.}~\bibnamefont
  {Boyd}},\ }\bibfield  {title} {\bibinfo {title} {A rewriting system for
  convex optimization problems},\ }\href@noop {} {\bibfield  {journal}
  {\bibinfo  {journal} {Journal of Control and Decision}\ }\textbf {\bibinfo
  {volume} {5}},\ \bibinfo {pages} {42} (\bibinfo {year} {2018})}\BibitemShut
  {NoStop}%
\bibitem [{\citenamefont {Schapeler}()}]{schapeler2020cvxpycode}%
  \BibitemOpen
  \bibfield  {author} {\bibinfo {author} {\bibfnamefont {T.}~\bibnamefont
  {Schapeler}},\ }\href@noop {} {\bibinfo {title} {Detector tomography python
  code}},\ \bibinfo {note}
  {\url{https://physik.uni-paderborn.de/fileadmin/physik/Arbeitsgruppen/bartley/Downloads/Detector-Tomography-CVXPY-Python-Timon-Schapeler.zip}}\BibitemShut
  {NoStop}%
\bibitem [{\citenamefont {Ramilli}\ \emph {et~al.}(2010)\citenamefont
  {Ramilli}, \citenamefont {Allevi}, \citenamefont {Chmill}, \citenamefont
  {Bondani}, \citenamefont {Caccia},\ and\ \citenamefont
  {Andreoni}}]{ramilli2010photon}%
  \BibitemOpen
  \bibfield  {author} {\bibinfo {author} {\bibfnamefont {M.}~\bibnamefont
  {Ramilli}}, \bibinfo {author} {\bibfnamefont {A.}~\bibnamefont {Allevi}},
  \bibinfo {author} {\bibfnamefont {V.}~\bibnamefont {Chmill}}, \bibinfo
  {author} {\bibfnamefont {M.}~\bibnamefont {Bondani}}, \bibinfo {author}
  {\bibfnamefont {M.}~\bibnamefont {Caccia}},\ and\ \bibinfo {author}
  {\bibfnamefont {A.}~\bibnamefont {Andreoni}},\ }\bibfield  {title} {\bibinfo
  {title} {Photon-number statistics with silicon photomultipliers},\
  }\href@noop {} {\bibfield  {journal} {\bibinfo  {journal} {Journal of the
  Optical Society of America B}\ }\textbf {\bibinfo {volume} {27}},\ \bibinfo
  {pages} {852} (\bibinfo {year} {2010})}\BibitemShut {NoStop}%
\bibitem [{\citenamefont {Kalashnikov}\ \emph {et~al.}(2011)\citenamefont
  {Kalashnikov}, \citenamefont {Tan}, \citenamefont {Chekhova},\ and\
  \citenamefont {Krivitsky}}]{kalashnikov2011accessing}%
  \BibitemOpen
  \bibfield  {author} {\bibinfo {author} {\bibfnamefont {D.~A.}\ \bibnamefont
  {Kalashnikov}}, \bibinfo {author} {\bibfnamefont {S.~H.}\ \bibnamefont
  {Tan}}, \bibinfo {author} {\bibfnamefont {M.~V.}\ \bibnamefont {Chekhova}},\
  and\ \bibinfo {author} {\bibfnamefont {L.~A.}\ \bibnamefont {Krivitsky}},\
  }\bibfield  {title} {\bibinfo {title} {Accessing photon bunching with a
  photon number resolving multi-pixel detector},\ }\href@noop {} {\bibfield
  {journal} {\bibinfo  {journal} {Optics Express}\ }\textbf {\bibinfo {volume}
  {19}},\ \bibinfo {pages} {9352} (\bibinfo {year} {2011})}\BibitemShut
  {NoStop}%
\bibitem [{\citenamefont {Dovrat}\ \emph {et~al.}(2012)\citenamefont {Dovrat},
  \citenamefont {Bakstein}, \citenamefont {Istrati}, \citenamefont {Shaham},\
  and\ \citenamefont {Eisenberg}}]{dovrat2012measurements}%
  \BibitemOpen
  \bibfield  {author} {\bibinfo {author} {\bibfnamefont {L.}~\bibnamefont
  {Dovrat}}, \bibinfo {author} {\bibfnamefont {M.}~\bibnamefont {Bakstein}},
  \bibinfo {author} {\bibfnamefont {D.}~\bibnamefont {Istrati}}, \bibinfo
  {author} {\bibfnamefont {A.}~\bibnamefont {Shaham}},\ and\ \bibinfo {author}
  {\bibfnamefont {H.}~\bibnamefont {Eisenberg}},\ }\bibfield  {title} {\bibinfo
  {title} {Measurements of the dependence of the photon-number distribution on
  the number of modes in parametric down-conversion},\ }\href@noop {}
  {\bibfield  {journal} {\bibinfo  {journal} {Optics Express}\ }\textbf
  {\bibinfo {volume} {20}},\ \bibinfo {pages} {2266} (\bibinfo {year}
  {2012})}\BibitemShut {NoStop}%
\end{thebibliography}%

\end{document}